\newif\ifpnas
\pnasfalse 
\ifpnas
    \documentclass[9pt,twocolumn,twoside]{pnas-new}
    \usepackage{bm}
\else
  \documentclass[twocolumn,reprint,floatfix]{revtex4-1}
  \usepackage[utf8]{inputenc}
  \usepackage{newtxtext}
  \RequirePackage[dvipsnames,usenames]{xcolor}
  \usepackage{hyperref}
\usepackage{bm}

  \usepackage{color} 
  \usepackage{amsmath} 
  \usepackage{graphicx} 

  \usepackage[english]{babel}
\fi

\ifpnas
    \templatetype{pnasresearcharticle} 
\else
  \makeatother
\begin{document}
\fi
\title{Mean field theory of self-organizing memristive connectomes}
\ifpnas

\author[a,1]{F. Caravelli}
\author[b,1,2]{G. Milano} 
\author[c]{C. Ricciardi}
\author[d]{Z. Kuncic}

\affil[a]{Theoretical Division (T4),
Los Alamos National Laboratory, Los Alamos, New Mexico 87545, USA}
\affil[b]{Advanced Materials Metrology and Life Sciences Division, INRiM (Istituto Nazionale di Ricerca Metrologica), Strada delle Cacce 91, 10135 Torino, Italy}
\affil[c]{Department of Applied Science and Technology, Politecnico di Torino, C.so Duca degli Abruzzi 24, 10129 Torino, Italy}
\affil[d]{School of Physics, University of Sydney, Sydney, NSW 2006, Australia}

\leadauthor{Caravelli}

\significancestatement{

In the connectome of self-organizing memristive nanonetworks, memristive elements are interconnected in a specific pattern to form a complex network. Here, the emergent behavior arise from the mutual interaction in between its constitutive components similarly to biological neuronal circuits. Besides applications in flexible electronics, solar cells, sensors and energy storage devices, nanowire networks have been proposed as platforms for hardware realization of neural networks making possible the \textit{in materia} implementation of unconventional computing paradigms such as reservoir computing. In this context, the junctions between intersecting nanowires are well described as memristive elements, which introduce  memory effects into the system and are responsible for the emergent neuromorphic behavior of the system.
In this work, we make a fundamental step towards elucidating the collective dynamics of nanowire connectomes, obtaining a model to describe the behavior of a large number of nanowires, and that explains the collective behavior of the connectome in response to electrical stimuli.  
}

\authorcontributions{F. Caravelli derived the equations and wrote the initial draft. G. Milano performed the experiments. All authors contributed to the study conception and design, and to the writing of the manuscript.}
\authordeclaration{No conflict of interest.}
\correspondingauthor{\textsuperscript{2}To whom correspondence should be addressed. E-mail: caravelli@lanl.gov}

\keywords{random silver nanowires $|$ memristors $|$ mean field theory $|$ conductance}

\else
\author{Francesco Caravelli\textsuperscript{1}, Gianluca Milano\textsuperscript{2}, Carlo Ricciardi\textsuperscript{3}, Zdenka Kuncic\textsuperscript{4}}

  \affiliation{
  $^1$ Theoretical Division (T-4), Los Alamos National Laboratory, Los Alamos, New Mexico 87545, USA\\
  $^2$ Advanced Materials Metrology and Life Sciences Division, INRiM (Istituto Nazionale di Ricerca Metrologica), Strada delle Cacce 91, 10135 Torino, Italy\\
  $^3$
  Department of Applied Science and Technology, Politecnico di Torino, C.so Duca degli Abruzzi 24, 10129 Torino, Italy\\
  $^4$School of Physics, University of Sydney, Sydney, NSW 2006, Australia}

\fi

\newcommand{\zk}[1]{{\textcolor{red}{Zdenka: #1}}}
\newcommand{\gm}[1]{{\textcolor{blue}{Gianluca: #1}}}
\newcommand{\crd}[1]{{\textcolor{purple}{Carlo:#1}}}
\newcommand{\fc}[1]{{\textcolor{blue}{Francesco:#1}}}

\begin{abstract}

Biological neuronal networks are characterized by nonlinear interactions and complex connectivity.  Given the growing impetus to build neuromorphic computers, understanding physical devices that exhibit structures and functionalities similar to biological neural networks is an important step toward this goal.

 Self-organizing circuits of nanodevices 
are at the forefront of the research in neuromorphic computing, as their behavior mimics synaptic plasticity features of biological neuronal circuits. However, an effective theory to describe their behavior is lacking. 

This study provides for the first time an effective mean field theory for the emergent voltage-induced polymorphism of \textit{circuits} of a nanowire connectome, showing that the behavior of these circuits can be explained by a low-dimensional dynamical equation. The equation can be derived from the microscopic dynamics of a single memristive junction in analytical form. We test our effective model on experiments of nanowire networks
and show that it fits both the potentiation and depression of these synapse-mimicking circuits. We show that our theory applies beyond the case of nanowire networks by formulating a general mean-field theory of conductance transitions in self-organizing memristive connectomes.
\end{abstract}

\ifpnas
  \dates{This manuscript was compiled on \today}
  \doi{\url{www.pnas.org/cgi/doi/10.1073/pnas.XXXXXXXXXX}}

  \begin{document}

  \maketitle
  \thispagestyle{firststyle}
  \ifthenelse{\boolean{shortarticle}}{\ifthenelse{\boolean{singlecolumn}}{\abscontentformatted}{\abscontent}}{}





\else
  \maketitle
  \section*{Introduction}

\fi

\ifpnas  \else  \fi 


Unconventional physical systems consisting of many interacting components have been proposed for the realization of self-organizing and biologically plausible behavior where the response to electrical stimuli mimics features typical of neuronal circuits \cite{Zdenkaadvphys}.

\begin{figure}[ht!]
    \centering
\includegraphics[scale=0.35]{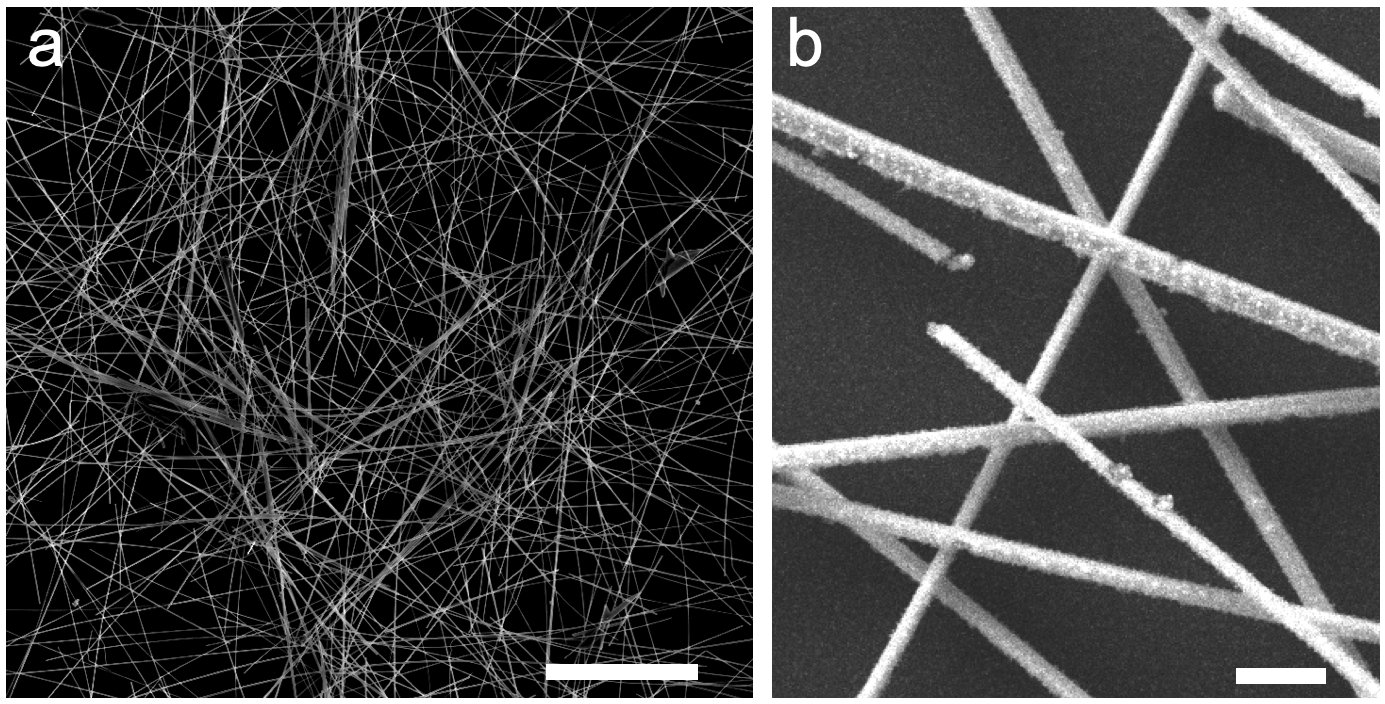}
    \caption{Connectome of a representative self-organizing nanowire network. (a) SEM image of a network of highly interconnected Ag nanowires (scale bar, 10$\mu$m); (b) magnified area showing nanoscale cross-point junctions between intersecting NWs (scale bar, 400 nm).}
    \label{fig:junctions}
\end{figure}

Metallic nanowire (NW) networks are self-assembled networks of interconnected NWs that can be used for various applications, such as in electronics \cite{elect}, energy storage \cite{enstor}, sensors \cite{sens} and machine learning \cite{milanores}. Among metallic networks, silver (Ag) NW networks have attracted great attention for the realization of neuromorphic devices and architectures \cite{Zdenkaadvphys,nakayama, avizienis, milanores}. 
Self-assemblies of NWs are intriguing 
complex physical systems \cite{infprop}, formed by randomly dispersing NWs with diameter in the order of tens of nanometers on a substrate. A self-assembled Ag-NW network is shown in Fig. \ref{fig:junctions}\textit{(a)}. It is evident that these form intricate patterns of connectivity. Despite the apparent complexity of these networks, models for the generation of these networks mimicking the realistic formation of the NW network have been proposed in the literature \cite{infprop,hochstetteretal,MILANO2022137}, reproducing the almost two-dimensional structure of the circuit, and local properties such as average degree.
Additionally, the intersection between two NWs (as shown in Fig. \ref{fig:junctions}\textit{(b)}) act as electrical junctions with all the nonlinear characteristics of a memristive component \cite{Milano2019,Manning2018,milano2020,memr1,memr2,yang}, making these systems promising platforms for the realization of neuromorphic electronic systems \cite{mead}. 


A memristive component is a one-port (two-terminal) device where the internal resistance state depends on the history of applied voltage or current \cite{stru08}.
Two-terminal memristive devices are considered fundamental
building blocks for the physical realization of artificial neural networks \cite{Zhang2020}. Memristors act as  artificial synapses,
and over the last few years, both ordered and disordered circuits of memristors have been studied theoretically and experimentally in the literature. Ordered networks of memristive devices, arranged in arrays
of  conventional crossbar architecture, have been used in a variety of supervised and unsupervised machine learning tasks, showing that these are apt for the implementation of brain-inspired computational frameworks \cite{Xia2019}. However, it has been suggested in the literature that brain-inspired computation can also be achieved in disordered networks of memristive devices \cite{reviewCarCar}. 
In particular, memristive devices can also be implemented using litographically printed magnetic nanoislands, both in order and disordered arrays \cite{saccone1,caravelli_2022}, and tailored for computational purposes  \cite{caravelli_2020,gartside} .

 In particular, memristive elements in these networks can endow short-term synaptic plasticity that is related to internal dynamics of memristive components \cite{plast,AtomicSwitch1,AtomicSwitch2,milanor}, making memristive NW networks suitable platforms for \textit{in materia} implementation of reservoir computing \cite{sheldon,milano2020}.
 The wiring diagram of a large number of memristive nanowires forms an artificial \textit{connectome}, e.g. a network of nanowires and junctions. However, it is still unclear how collective dynamics and synaptic functionalities emerge coherently from such a complex connectome.
As we show in this study, this is a property of memristive components arranged on a complex network.

The graph statistical properties properties of a connectome (such as the local number of connections) of NW network models have been studied in \cite{loeffler2020,MILANO2022137} together with emerging memristive dynamics \cite{infprop}, providing a quantitative agreement with the existing experimental results within the context of Ag NW networks.  
The resistivity of these networks is mainly due to the voltage drop at the junctions (since $G_{jun}\ll G_{wire}$, \cite{milano2020}).
This means that as a first approximation, one can neglect the resistivity of the wires and consider a network of ideal memristive junctions, whose behavior has to be then carefully analyzed. In the equivalent circuit, such approximation implies that Ag nanowires become effective nodes of the circuit, while junctions become memristive links.
The key aspect of the present study is that the transition between low and high network conductance states can be described by a mean-field theory.

\section*{Background}
The interesting properties of Ag NW networks have been probed experimentally over the last decade \cite{avizienis,nakayama}.  The conductivity of the single junction and networks has been studied in detail, and its behavior emerges from the interplay of roughly two effects, depending on the composition of the wires.
First, there are many geometrical effects, due to the distribution of the wires, which are not the scope of the present study. For instance, at low density of nanowires, there are few or no percolating paths between two nodes where the probes are attached. We are interested in the dynamic effects of conductivity, in particular, transitions between low and high conductance states \cite{nakayama,hochstetteretal}. 
In \cite{caravelliscience}, conductance transitions were predicted for circuits composed of the simplest type of memristive devices using the Strukov-Williams model for TiO$_2$ memristors \cite{stru08,stru}, which is a bulk model for filament conductance. The dynamical component is due to metal filament formation across the junction, due to the voltage-induced migration of Ag$^+$. Moreover, quantum tunneling also introduces a source of nonlinearity, but this becomes important only for nearly ungapped filaments \cite{hochstetteretal}. 

The internal dynamics of NW junction memristive elements characterized by short-term memory can be described by a rate--balance equation \cite{Mirandeetal}. This is a dynamical model that can be used to describe the conductivity of the NW junction exhibiting a nonlinear dynamical response to a voltage bias, due to the formation of a metallic filament.
For the type of Ag NW network experiments that we are interested in,  the effective model which well describes  the conductance of a single junction is a rate equation \cite{MILANO2022137,Mirandeetal}.
This dynamical model for the junction conductivity depends on two  parameters, $G_{min}$ and $G_{max}$, representing the minimum and maximum conductance, and voltage-drop dependent rate constant $\eta_P$ and $\eta_D$:

\begin{eqnarray}
    G(g)&=&G_{min} g+G_{max}(1-g)=G_{min}(1+\chi g) \label{eq:conddyn0}\\
    \frac{dg}{dt}
    &=&\eta_{P}(\Delta v)(1-g)-\eta_{D}(\Delta v)g,
    \label{eq:conddyn}\\
    \eta_{P}(\Delta v)&=&\kappa_{P0}\  \text{exp}(\eta_{P0} \Delta v),  \label{eq:conddyn1s} \\
    \eta_{D}(\Delta v)&=&\kappa_{D0} \ \text{exp}(-\eta_{D0} \Delta v) \label{eq:conddyn2s}
\end{eqnarray}
Above, $G(g)$ is the junction conductance, $g$ is the normalized conductance with $0\leq g\leq 1$, $\Delta v$ is the voltage drop on the junction. We have also introduced $\chi=(G_{max}-G_{min})/G_{min}$, which can be interpreted as the degree to which the system presents memory effects. In fact, if $\chi=0$, then these memristive elements become simple resistors. The parameter $\chi$ not only introduces then the nonlinearity in the system, but also induces the extent to which the system remembers the past states.
Of course, in a circuit, the behavior of the conductance of the single junction is contained in the voltage drop $\Delta v$, and thus through the graph representing the circuit. In what follows, since we will have $N$ junctions, we will refer to ${\bm G}(\vec g)$ as the diagonal matrix of the conductances, and $g_i$ the normalized conductance of the $i$-th junction. The voltage drops $\Delta v$ are generalized to a vector accordingly.

Disordered circuits such as those emerging in self-assembling nanowires present a variety of phenomena, and their architecture is closer to biological neuronal networks \cite{hochstetteretal,MILANO2022137}. However, the fact that experimentally a rather similar behavior is observed in many  differently self-assembled nano-structures suggests the existence of an underlying mechanism, such as self-averaging, explaining such homogeneity in responses.
Overall, the underlying complexity stands in the combination of spatiotemporal disorder, the nonlinear memory property of the single junction (cf. \eqref{eq:conddyn}), and the induced correlations between the junctions. In such a circuit, one has also to solve Kirchhoff laws. Let us call $\mathcal G$ the \textit{directed} graph representing the circuit, with edges oriented according to the positive currents $\vec i$ in the junctions. 
The directionality is indeed artificial, so if the direction of the edge was chosen to be say $+$, then a negative current means a current going in the opposite orientation as chosen. If the current is zero, there is no inconsistency in this case, because $0$ is the only number $m\in \mathbf{R}$ such that $-m=m$.
We call $B$ the directed incidence matrix of $\mathcal G$. Then, if the circuit is controlled by injecting a current between two nodes $n_1,n_2$, e.g. $+|j|$ at $n_1$ and $-|j|$ at $n_2$, the Kirchhoff laws can be obtained by solving the nodal analysis equations
\begin{eqnarray}
B\vec i&=&\vec j_{ext}\label{eq:kirchh0}\\
{\bm G} \Delta \vec v&=&\vec i, \label{eq:kirchh1}
\end{eqnarray}
where $(j_{ext})_i=0$ if $i\neq n_1,n_2$ and $(j_{ext})_i=\pm j$ for $i=n_1,n_2$ respectively. We see then that in order to simulate a circuit of $N$ junctions, we need to solve $N$ dynamical equations from \eqref{eq:conddyn} and $2N$ linear equations from \eqref{eq:kirchh1}, and finally, calculate the effective conductance. Here, we show that these $3N$ equations can be reduced to a single mean-field equation in which the parameters and voltages in \eqref{eq:conddyn0}-\eqref{eq:conddyn2s} are renormalized. In particular, we  derive the effective conductance $G_{eff}$ between nodes $n_1$ and $n_2$.

\begin{figure}
    \centering
    \includegraphics[scale=0.3]{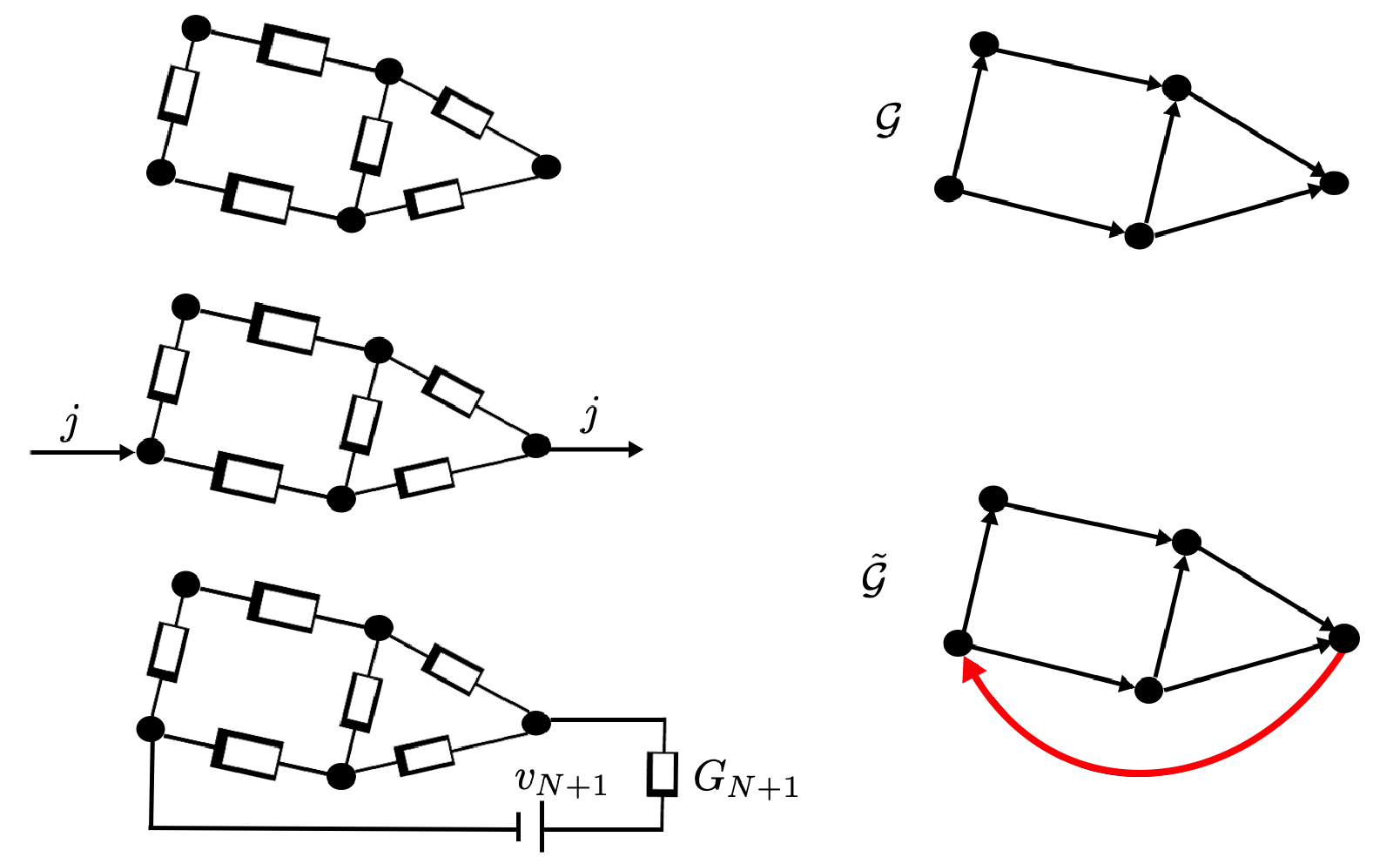}
    \caption{Effective graphs for the core circuit (top row, left) and effective graph (top row, right) and the effective circuit with currents (middle row) and with a voltage generator (bottom row). For any external generator (either current or voltage), there is an extra edge, which is highlighted in red in $\mathcal G$ on the right. In this work, we consider only a single extra edge.}
    \label{fig:effectiveg}
\end{figure}

\section*{Results}
\subsection*{Key formulae and mean-field theory}

One of the key advancements of this study is a technical intermediate step that allows integrating analytically the Kirchhoff laws of  \eqref{eq:kirchh0}-\eqref{eq:kirchh1}.
In order to derive a mean-field equation, we use a graph theoretical formalism to formally evaluate Kirchhoff's laws. We call $\tilde {\mathcal G}=\mathcal G\cup \mathcal G_{n_1n_2}$, e.g. the graph with the single edge $\mathcal G_{n_1n_2}=(n_1,n_2)$ added to the graph (see Fig. \ref{fig:effectiveg}). Formally, it possible to convert the current $\vec j_{ext}$ into an external voltage source $v_{N+1}$, in series to a conductance $G_{N+1}$, and satisfying $G_{N+1} v_{N+1}=j$. Since we now have $N+1$ voltage drops in the circuit, the vector of all voltage drops is given by $\Delta v_{all}$, where the first $N$ components are $\Delta \vec v$ and the $(N+1)$-th component is the voltage drop on the voltage source branch. As we show in the Supplementary Information (Sec. A), we can write $\vec j_{ext}=G_{N+1} \tilde B \vec v_s$, where $(\vec v_s)_i=0$ for $i\neq n_1,n_2$ and $(\vec v_s)_i=\pm v_{N+1}/2$ for $i=n_1,n_2$  respectively.
The parameter $g_{N+1}$ must satisfy $G(g_{N+1})=G_{N+1}$, and it can be shown that $\lim_{G_{N+1}\rightarrow 0} g_{N+1}=-\frac{1}{\chi}$.
 
Let $\tilde {\mathcal G}$ be the augmented graph with an extra directed edge between node $n_1$ and $n_2$, and $\tilde B$ the corresponding directed incidence matrix. 
As shown below, the voltage drops can be found analytically, thus avoiding solving numerically for \eqref{eq:kirchh0}-\eqref{eq:kirchh1}. In fact, we have (See Supplementary Information A1):\\
\textbf{Lemma 1 - Network voltage integration}:
For a circuit composed of memristive junctions satisfying \eqref{eq:conddyn0}, we have the following identity
\begin{eqnarray}
  \Delta \vec v_{all}=\lim_{G_{N+1}\rightarrow 0} \frac{G_{N+1}}{G_{min}} (I+\chi  \Omega {\bm g})^{-1} \Omega \vec v_s
  \label{eq:kirchhoff}
\end{eqnarray}
where $ \Omega=\tilde B^t (\tilde B\tilde B^t)^{-1}\tilde B$ is a projector operator and ${\bm g}$ is the diagonal matrix $\text{diag}(\vec g,g_{N+1})$. The relevance of \eqref{eq:kirchhoff} is that Kirchhoff's laws have been integrated analytically.  The underlying physical reason for the introduction of a projector operator, which has the property $\Omega^2=\Omega$,  is that it enforces the conservation of currents at the nodes \cite{Caravelli2019}. The matrix $\tilde B$ is the directed incidence matrix of the augmented graph $\tilde {\mathcal G}$ of Fig. \ref{fig:effectiveg} (bottom right), in which we have added an extra edge where either the voltage or current generator has been added. The matrix $\Omega$, which represents the interactions between elements due to Kirchhoff's laws, has been studied in detail previously \cite{caravelli2016rl}. In the case of planar circuits, for example, it was found that the interactions fall off exponentially with distance \cite{Caravelli2017}.

It is important to mention that \eqref{eq:kirchhoff} is useful since we can obtain both the voltage drops for the single junctions, which we can insert in \eqref{eq:conddyn} to time evolve the conductances and to obtain the effective behavior of the memristive network. In fact, the $N+1$ component of \eqref{eq:kirchhoff} must satisfy the equation $\Delta v_{N+1}G_{eff}=j$. Thus, one has to separate the matrix inverse of \eqref{eq:kirchhoff} in blocks. 
In order to do that, first we divide $\Omega$ on the subgraphs $\mathcal G$ and $\mathcal G_{n_1n_2}$ as
\begin{equation}
    \Omega=
\begin{pmatrix}
\tilde \Omega & \vec \Omega\\
\vec \Omega^t & \Omega_{N+1}
\end{pmatrix}\label{eq:sepmat}
\end{equation}
Let us define the quantity
\begin{equation}
\rho= \Omega_{N+1}- \chi\vec \Omega^{t}\tilde{\bm{g}} (I+\chi \tilde \Omega \tilde{\bm{g}})^{-1} \vec \Omega,
\label{eq:eta}
\end{equation}
where ${\bm \tilde g}=\text{diag}(\vec g)$, e.g. the parameters associated with the junctions. Let us now give a physical interpretation of these two quantities.

From the definitions above 
, we proved the following Corollaries (See Supplementary Information A, Sec. 4 and Sec. 6):\\
\textbf{Corollary 1 - Voltage drops}:
Let $\tilde G$ be an augmented circuit composed of memristive junctions
of the form of \eqref{eq:conddyn0}. Then the voltage drops on the junctions are given by
\begin{eqnarray}
\Delta \vec v
&=&\frac{ v_{N+1}}{1-\rho}\frac{G_{N+1}}{G_{min}}
(I+\chi \tilde \Omega \tilde{\bm{g}})^{-1}\vec \Omega.
\label{eq:vexactmt}
\end{eqnarray}
We can also extract the effective conductance, and we have\\
\textbf{Corollary 2 - Effective conductance}:
Let $\tilde G$ be an augmented circuit composed of memristive junctions 
of the form of \eqref{eq:conddyn0}. Then the effective conductance between node $n_1$ and $n_2$ is given by
\begin{eqnarray}
G_{eff}&=&G_{min}\frac{1+\chi g_{N+1}\rho}{\rho}\label{eq:effgeff}
\end{eqnarray}
We see that Corollary 1 and Corollary 2 are formal statements regarding the voltage drops and effective resistance as a function of the parameters $\vec g$ and the circuit topology, contained in $\Omega$. The Lemma and Corollaries above can also be generalized to nonlinear conductance functions, and we will see an example below and in Supplementary Information.

Let us provide a brief interpretation of \eqref{eq:vexactmt}. 
The vector $\vec \Omega$ can be thought of as a network backbone of the response function, e.g. the effective voltage on junction $i$ must be proportional to $(\vec \Omega)_i$. Effectively, \eqref{eq:vexactmt} is the solution of the voltage integration across the network, starting from the assumption that the voltage is applied between two nodes, inducing the separation of the matrix $\Omega$ given in \eqref{eq:sepmat}. The matrix $\tilde \Omega$ enters instead in the matrix inverse multiplying the internal junction conductances. Instead, \eqref{eq:effgeff} is important as it provides an interpretation of the quantity $\rho$ defined in \eqref{eq:eta} in terms of global effective conductance.

However, these are static statements, which do not take into account the fact that the junction conductances change over time. To derive an effective mean-field theory, we introduce an effective mean-field variable $\langle g(t)\rangle$ for the junction conductances. 
\begin{figure*}
    \centering
    \includegraphics[width=\textwidth]{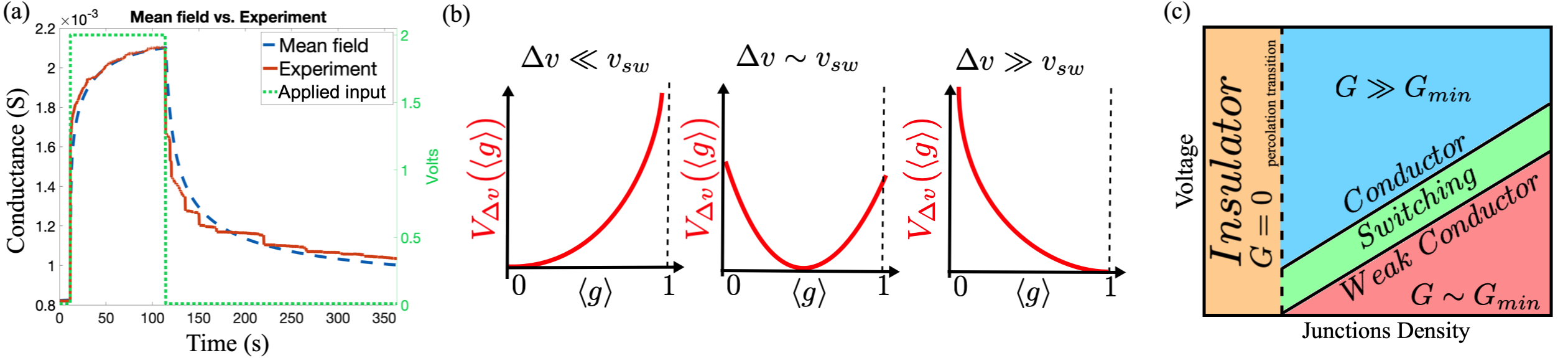}
    \caption{Effective mean-field behavior. \textit{(a)}  Conductance response to a 2~V square signal, followed by a low readout voltage, delivered to an Ag-PVP NW network device (green, experimental data from ref. \cite{milanores}) and mean-field theory fit to the experimental data (dashed blue). \textit{(b)} Switching of the effective potential $V_{\Delta v}$ 
    as a function of the order parameter $\langle g \rangle$ and applied voltage $\Delta v$. At low voltages, the system remains in a low conducting phase. At higher $\Delta v$, the potential switches and the system becomes conductive. This behavior can be inferred from eqn.~\eqref{eq:gmfe}. \textit{(c)} Schematic phase diagram of the system. At very low junction densities, the system does not have a conducting path. Above a percolation threshold, the system establishes a connected path and the conductance becomes a function of voltage. 
    To the right of the insulating region, the system undergoes a phase transition as it switches to a conducting phase for a given density of memristive elements on the shortest path across the bias. This phase diagram depends on the initial conditions of the junctions; here we assume them to be all in the low-conductance state (i.e. homogeneous system). The diagram can indeed change for other systems and different initial conditions of the junctions, but we expect the general structure to be preserved.
    }
    \label{fig:fitplotpd}
\end{figure*}

\textbf{Mean-field approximation}. All the equations above are exact. However, we can see that we are still left with a matrix inverse given by $(I+\chi \tilde \Omega \tilde{\bm{g}})^{-1}$. In order to simplify the equations and obtain a lower-dimensional system, we define the mean-field variable $\langle g\rangle$ via
\begin{eqnarray}
    \langle g\rangle=\text{argmin}_{\bar g}\|(I+\chi \tilde \Omega \tilde{\bm{g}})-(I+\bar g\chi \tilde \Omega) \|_F^2,
\end{eqnarray}
where $\|\|_F^2$ represents the Frobenius matrix norm-squared, i.e. $\|M\|_F^2 = \text{Tr}(M M^t)$. 
It is indeed easy to see that if it minimizes the function above, it also minimizes a similar definition with the matrix inverses. The exact solution is given by $\langle g\rangle=\frac{\text{Tr}(\tilde \Omega^2\bm{g})}{\text{Tr}(\tilde \Omega^2)}$. The result is thus a complex function defined in terms of the single junction parameters $\vec g$, and the network connectivity. This might seem at first a drawback, as the mean-field parameter we are interested in is defined in terms of a large number of unknown parameters, including the network topology. However, as we show below, if we assume that such a mean-field order parameter exists, we reduce the number of parameters to be fit experimentally to only four plus $\langle g\rangle$; these can then be fit experimentally. First, it can be shown that
\begin{eqnarray}
G_{eff}(\langle g\rangle)&=&\frac{1-\Omega_{N+1}}{\Omega_{N+1}}G(\langle g\rangle).
\end{eqnarray}
We thus have that depending on where the external voltage (or current) generator simply is reabsorbed into the $G_{min}$ and $G_{max}$ parameters, and the conductance parameters can be fit experimentally using the same model. The voltage for each memristive junction 
is given, in the mean-field approximation, by
\begin{eqnarray}
\Delta \vec v\approx \Delta \vec v_{mft}=\frac{G_{N+1} v_{N+1}}{G_{min}(1-\Omega_{N+1})} \frac{1}{1+\langle g\rangle \chi}\vec \Omega
\label{eq:vmftmt}
\end{eqnarray}
where the vector $\vec \Omega$ represents the response of each memristive element when a voltage is applied to the network between nodes $n_1$ and $n_N$. We now perform the second approximation. We replace $\vec \Omega$ with $\langle \Omega\rangle \vec 1$. Then, at this point summing cleverly on the left-hand side gives a self-consistent \textit{single} memristor equation (details in the Supplementary Information, Sec. C), in which the parameters $\eta_P$, $\eta_D$, $\kappa_P$, and $\kappa_D$ are multiplied by network-dependent quantities. The applied voltage is instead multiplied by a factor
\begin{equation}
    \Delta v\rightarrow \Delta v/(1+\chi^{eff} \langle g\rangle).
\end{equation}
We thus see that by putting all these intermediate results together, we do obtain an effective system of equations as those in \eqref{eq:conddyn0},\eqref{eq:conddyn}. 
This is the case in a typical experimental setup, in which the typical conductance measurement involves placing two probes between two (or more) nanowires. Thus, the quantity of interest is an effective resistance, which depends on the point of contact. Thus, our theory describes the effective conductance measurement of this complex network of nanowires, and the effective dynamical equations for the conductance are given by
\begin{eqnarray}
    &&G(\langle g\rangle)=G_{min}^{eff}(1+\chi^{eff} \langle g\rangle ) \label{eq:conddyneff0}\\
    &&\frac{d\langle g\rangle }{dt}
    =\eta_{P}^{eff}(\Delta v,\langle g\rangle)(1-\langle g\rangle)-\eta_{D}^{eff}(\Delta v,\langle g\rangle)\langle g\rangle,
    \label{eq:conddyneff}\\
    &&\eta_{P}(\Delta v,\langle g\rangle)=\kappa_{P0}^{eff} \text{exp}\Big(\eta_{P0}^{eff} \frac{\Delta v}{1+\chi^{eff} \langle g\rangle}\Big),\label{eq:conddyneff1} \\
    &&\eta_{D}(\Delta v,\langle g\rangle)=\kappa_{D0}^{eff} \text{exp}\Big(-\eta_{D0}^{eff} \frac{\Delta v}{1+\chi^{eff} \langle g\rangle}\Big)\label{eq:conddyneff2}.
\end{eqnarray}
Above, $\langle g\rangle$ is an effective dynamical conductance parameter, which can be obtained from the microscopic values $g_i$ of the single junctions. The specific expression for $\langle g\rangle$ in terms of the $g_i$ and the circuit topology is not important from an effective macroscopic system perspective, as it is nonetheless self-consistent with the measurement of an initial value of the effective conductance of the sample. For the purpose of context, we used the same analytical techniques introduced in \cite{caravelli2016rl,Caravelli2019,caravelliscience}.

It is important to note that the free parameters are of the same number as the ones for the single junction. The key difference is that now $\Delta v$ is renormalized by a factor given by $1+\chi^{eff}\langle g\rangle$; the other parameters are also renormalized by network-dependent quantities. Clearly, the equation above has the advantage that one uses a single rate equation for the entire NW network.

\subsection*{Experimental validation}

Our experimental results are based on measurements of a NW network device using two electrical probes \cite{MILANO2022137}.
Self-assembling NWs were realized by drop-casting Ag NWs in suspension
on a SiO$_2$ insulating substrate \cite{milais}. A high density of NW cross-point junctions (106 junctions/mm$^2$) was achieved, ensuring that the network is above the percolation threshold. 
Ag NWs were passivated by a coating of PVP of 1–2 nm thickness around the Ag core  \cite{milano2020,MILANO2022137}. 
PVP acts as a solid electrolyte for the junctions, as an electrochemical metallization induces a memristive behavior to the junction, characterized by the rate equation \eqref{eq:conddyn}. We then applied a square voltage of 2\;V for 100\;s, followed by a small voltage for measurement purposes, as shown in Fig. \ref{fig:fitplotpd}(a); using this protocol, we are measuring the short-term memory of the sample.


To see that the mean-field equation can fit the response of a real device, we consider the best-fit parameters that minimize, given the input voltage $\Delta v(t)$, the error $E_T=\Big(\frac{1}{T}\int _0^Tdt\Big( G^{exp}(t)-G_{eff}^{mft}(t)\big)\Big)^2$. As we can see from Fig.~\ref{fig:fitplotpd}(a) the mean-field theory reproduces the behavior of the network. Thus, it can be used to obtain, given the tuned parameters, the behavior of the nanowires as a function of the maximum voltage $\Delta v$ applied to the device. 

The advantage of using a mean-field equation such as \eqref{eq:conddyneff} is that, since it is one-dimensional, we can always express it in terms of an effective potential $V$
\begin{eqnarray}
  \frac{d\langle g\rangle}{dt}=-\frac{dV_{\Delta v}(\langle g\rangle)}{d\langle g\rangle},
\end{eqnarray}
i.e. it shows that there is an effective low dimensional dynamics driven by a voltage-dependent mean-field potential $V_{\Delta v}$. This approach was previously applied to study current-controlled memristive circuits in \cite{caravelliscience}, where a change in symmetry of the potential occurs as a function of applied voltage.

The effective potential can be obtained analytically by integrating \eqref{eq:conddyneff}, giving $dV_{\Delta v}(\langle g\rangle)=-\int d\langle g\rangle \frac{d\langle g\rangle}{dt}$  (the exact expression is provided in the Supplementary Information Sec. B2). Let us, however, report here the phenomenology of the potential change.
Using the effective parameters obtained from the fit in Fig. \ref{fig:fitplotpd}(a), we estimate that 
there is a threshold at which the potential switches and the system transitions from a low to a high conducting phase. The switching of the potential occurs at very small values of $\Delta v_{sw}^{th}\sim 2 \cdot 10^{-2}$ volts. However, since the gradient is very shallow and it increases as a function of the voltage, a noticeable change in the effective conductance occurs, within the time scale of the tens-hundred seconds, for $\Delta v_{sw}^{exp}\sim 0.9$ volts, which is consistent with the experimental timescale. The picture we obtain is then the one of Fig. \ref{fig:fitplotpd}(b), in which the potential changes its minimum abruptly, but continuously, e.g. the equilibrium value of the effective conductance $\langle G_{eff}\rangle^{eq}$ changes from $G_{min}$ to $G_{max}$ as a function of $\Delta v$. As we can see, the effective description provides a qualitative and quantitative prediction of the conductance transition.

\subsection*{Conductance transitions} As in the case of  current controlled memristor networks studied in
\cite{caravelliscience},  the effective potential can be calculated analytically via approximations. It ought to be noted, however, that there the potential switching takes a different form, and that unlike here, it is an unstable fixed point that moves as a function of the effective (average) current in the circuit. There, the system can have two stable fixed points at the same time. In our case instead, the system has always a single stable fixed point, which rapidly switches as a function of the applied voltages. 

Nevertheless, the overall picture which emerges in both cases is similar and is the one shown in Fig. \ref{fig:fitplotpd}(c), replacing current with voltage. For sufficiently high circuit density (characterized by the number of memristive junctions), the mean-field description suggests that the system is in a low conductance state, and for larger applied voltages (or currents), the system switches to a high conductive state. This picture is qualitatively similar to other types of nanowire networks \cite{hochstetteretal}, where it was found that threshold dynamics can lead to avalanches. These critical dynamics were also studied using mean-field theory in \cite{sheldonava}. 

It thus seems then that there is a general pattern emerging concerning nonlinear circuits with memory, e.g. memristive circuits. At low densities of memristors, given the effective conductance between two nodes, the system is in an insulating phase because of the \textit{geometric} features of the circuit. At higher densities, above a percolation threshold, the probability of establishing a conductance path between two nodes becomes macroscopically large, and would also occur in a resistor network. 
Our study is then concerned with the region to the right of this transition, where between a weak conducting and a conducting phase there is a switching region. Whilst the details of a such region depends on the type of memristor and initial conditions of the system, the results of this study (analytical) and \cite{milano2020} (numerical) for Ag nanowires, those of \cite{caravelli2016rl,caravelliscience} for current controlled memristors (analytical), and those of \cite{hochstetteretal} for atomic switch NW networks \cite{avizienis} (numerical), suggest that such a phase diagram is robust to the details of the nonlinearity.
this is because for low nodes the current flows on a smaller number of junctions, thus having a larger voltage drop on each, thus making them switch earlier.

It is important to stress that the mean-field theory presented in this study is a result of the symmetries induced by Kirchhoff's laws, and that can be applied to a variety of other systems.

\subsection*{Other memristive systems}
To see the broad applicability of this mean-field technique, we provide the equations  for other models describing the dynamics of 
different self-organizing memristive networks.
These conductance transitions occur beyond a particular model; for this purpose, we use a model describing the behavior of both percolating nanoparticles \cite{BrownPart} and 3D nanowires \cite{DANIELS2022122}, but still constrained by the Kirchhoff laws. 



We consider the following model for the conductance $G(z)$ of each  junction, given by the set of equations
\begin{eqnarray}
  \frac{dz}{dt}=\mu\frac{V}{D-z}-\kappa z, G(z)=\alpha e^{-\beta (D-z)},
\end{eqnarray}
with $\mu= 0.346\ nm^2\ V^{-1}$ and $\kappa=0.038 s^{-1}$. $D$ is the distance between the nanoparticles or nanowires (in nm) and $z(t)$ represents the effective gap between the evolving nano-filament and the nano-wire/-particle. 
In the Supplementary Information (in Sec. B) we have obtained a generalization of Lemma 1 and the subsequent corollaries to the case of a junction whose conductance is not a linear function of the internal memory parameter $g$.

\begin{figure}
    \centering
    \includegraphics[scale=0.5]{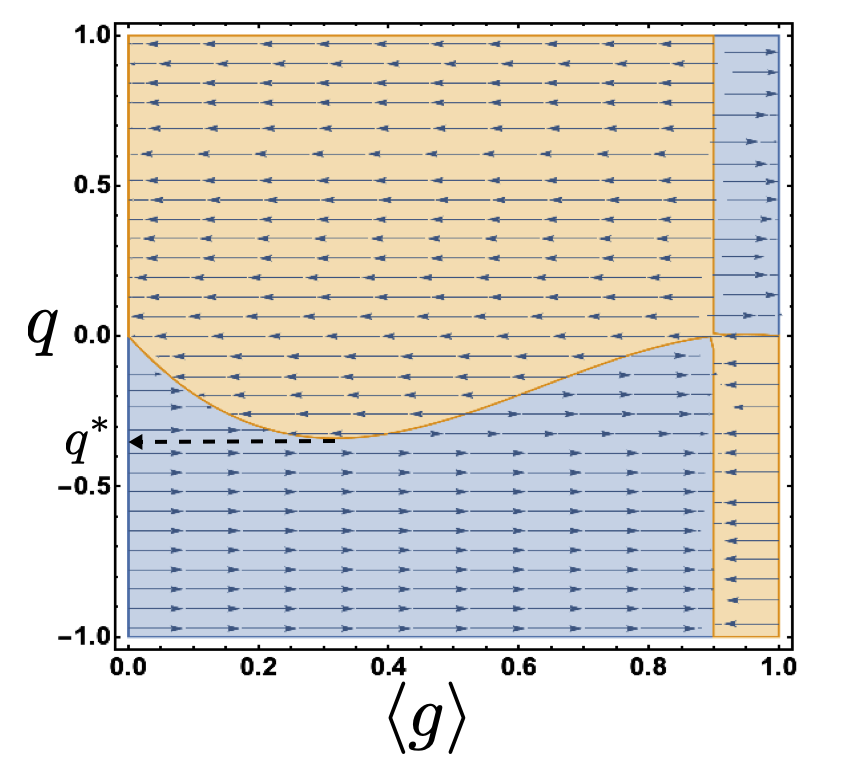}
    \caption{Effective force $(-\partial_z V(z))$ for the mean-field nano-particle/-wire filament formation, %
    for $\kappa=a=1$ as a function of the effective voltage $q$. For $q>0$, the model is most likely in the low conductance state, but for values $q<q^*$ we observe an abrupt transition from a low to a high conductance state.
    }
    \label{fig:regionplot}
\end{figure}

We provide here the necessary background to understand the model of \cite{DANIELS2022122}. Similarly to what we had done before, we rewrite the equations above in terms of a single parameter $g=z/D$.
Then, the effective mean-field can be obtained by imposing
$\vec g=\langle g\rangle \vec 1$. and we obtain the effective equations (see Supplementary Information Sec. B) 
\begin{eqnarray}
\frac{d}{dt}\langle g\rangle&=& \frac{q_{eff}}{\big(1-\langle g\rangle\big)\big(1+\chi_{eff} f(\langle g\rangle)\big)}-\kappa_{eff} \langle g\rangle\\
G_{eff}(\langle g\rangle) &=&G_{min}^{eff}\ (1+\chi_{eff}f(\langle g\rangle)).
\end{eqnarray}
with $f(x)=e^{-a(1-x)}-1$ and $\chi_{eff}=(G_{max}^{eff}-G_{min}^{eff})/G_{min}^{eff}$. 
Above, $q_{eff}=a v$, where $a$ is a proportionality constant depending on the microscopic parameters, while $v$ is an effective voltage.

With these equations in hand, we can see why the conductance transitions are not a feature of a particular model. An analysis of these equations shows that there is a first-order transition between a high and low conductance state. This can be seen in Fig. \ref{fig:regionplot}, where we plot the effective direction of the force. As we can see, from the mean-field theory of this model we predict a first-order transition as a function of the effective voltage. This is the same phenomenon observed in \cite{brownxx,hochstetteretal} for a similar type of nanowire network. As we explain below, we contend this is a robust phenomenon that goes beyond the specific details of the model, and that can be characterized by an effective theory \textit{a l\'a Landau}.

\section*{Effective theory of conductance transitions}
To understand when  and how these conductance transitions occur, let us focus on the equilibrium obtained mean field equation for memristive nanowires, given by the solution of the equation (see Supplementary Information Sec. B1)
\begin{equation}
    \langle g\rangle^*=\Big(1+ s e^{\frac{f_0 v}{1+\chi^{eff}} \langle g\rangle^*}\Big)^{-1}
    \label{eq:gmfe}
\end{equation}
where the parameters $f_0, s$ and $\chi^{eff}$ can be determined experimentally, but have an explicit form from the mean-field theory in terms of the microscopic parameters. 
It can be seen explicitly from the equilibrium
how the switching of Fig. \ref{fig:fitplotpd} (b) occurs as a function of the applied voltage $v$.

\begin{figure}
\includegraphics[scale=0.4]{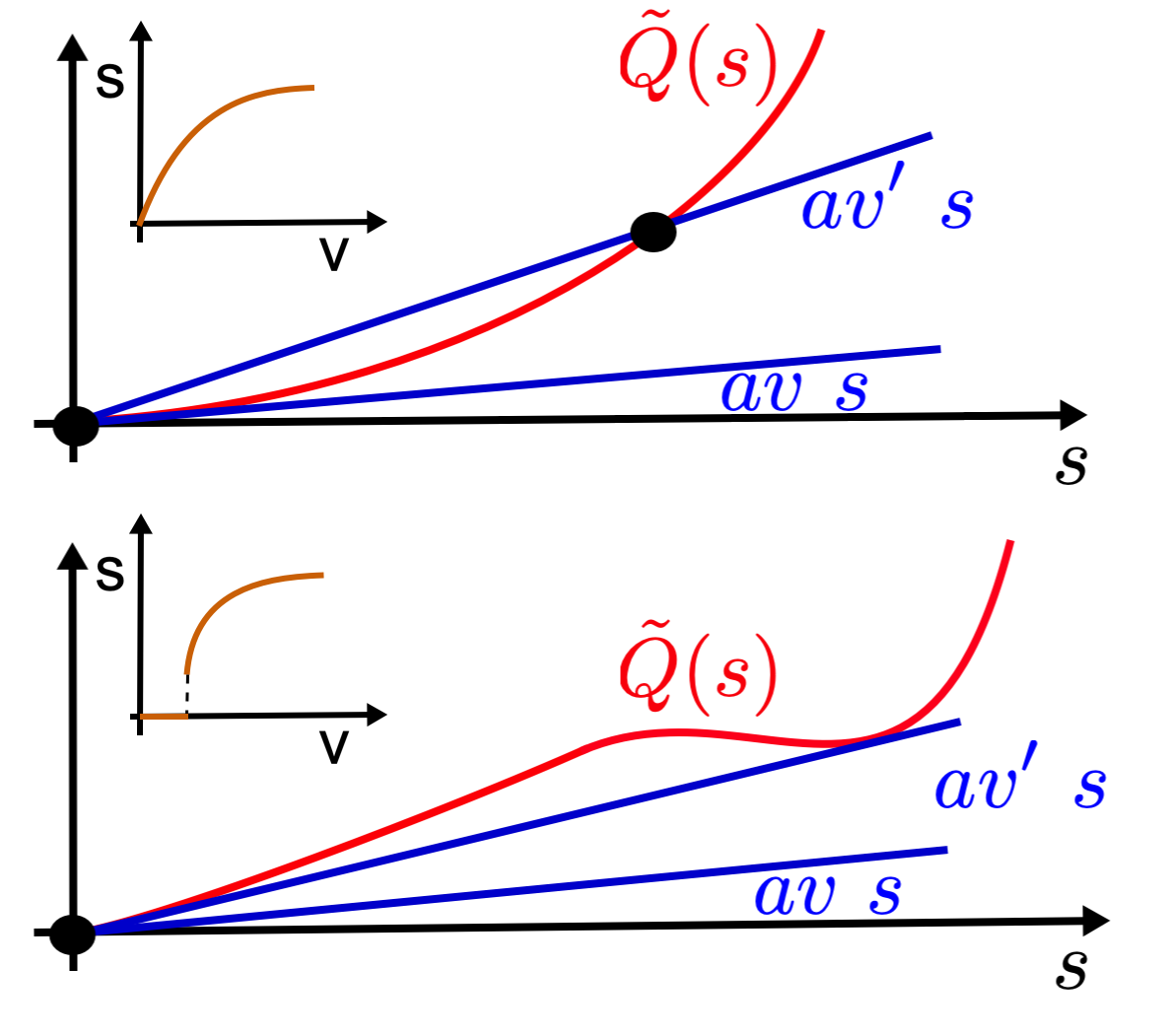}
\caption{Graphical representation of \eqref{eq:landau}. \textit{Top}. For small values of the applied voltage $v$, because of the convexity of $Q(\bar r)$ in $0$, the only mean field parameter allowed equilibrium point is $\bar r=0$, but as the voltage reaches a critical threshold, the system's effective order parameter $s$ (e.g. equilibrium conductance) varies smoothly as a function of voltage (inset). \textit{Bottom}. For a non-convex function $\tilde Q(x)$, we can have first-order transitions as a function of the voltage (inset).}
\label{fig:meanq}
\end{figure}

For small values of $v$, the effective mean field potential takes the form
\begin{equation}
    V(\langle g\rangle)=a \langle g\rangle + b\langle g\rangle^2-c v \log (1+\chi \langle g\rangle).
\end{equation}
where $a,b$ and $c$ are constants. In the case of the nanoparticles, such a potential can be written in the form
\begin{equation}
    V(\langle g\rangle)=a \langle g\rangle^2-b v \log (1+f(\langle g\rangle)).
\end{equation}
Similarly, for a network of memristors which satisfy $R(x)=R_{on} x+(1-x)R_{off}$ and $dx/dt=-\alpha x+i/\beta$,  the effective potential for the equivalent parameter $\langle x\rangle$ is given by
\cite{caravelliscience}
\begin{eqnarray}
    V(\langle x\rangle)=\frac{\alpha}{2} \langle x\rangle^2+\frac{v}{\chi}\log(1-\chi \langle x\rangle).
\end{eqnarray}
In all these cases which can be studied analytically, we thus see that the general form of the potential is written in the form
\begin{equation}
    V(\bar r)=\pm|Q(\bar r)-a v \log\big(1+P(\bar r)\big)|
\end{equation}
where $\bar r$ is a generic order parameter, and $Q(\cdot)$ and $P(\cdot)$ are generic functions, such that $Q(0)=Q^\prime(0)=P(0)=0$, i.e. there are no constant terms and for $v=0$ the only solution is $\bar r=0$.
The equilibrium points are then determined by the mean-field equation
\begin{eqnarray}
    \frac{\partial_{\bar r}Q(\bar r)}{\partial_{ \bar r} \log \big(1+P(\bar r)\big)}= a v.
\end{eqnarray}
If $P(\bar r)$ is a monotonic function, we can always define the effective order parameter given by $s=\log (1+P(\bar r))$ and then rewrite the expression above as the mean-field theory
\begin{equation}
    Q(\bar r(s))=\tilde Q(s)=a v s.
    \label{eq:landau}
\end{equation}
Using this formulation we see that the number of equilibrium points can be defined, as a function of the effective voltage $v$, \textit{a l\'a Landau}, depending on the function $\tilde Q(s)$. For small values of $v$, there is only one fixed point $s=0$, corresponding to the mean field parameter $\bar r=0$. For larger values, depending on the function $\tilde Q(s)$, there can be multiple fixed points. However, if the function $\tilde Q(s)$ is globally convex, the transition is continuous, which is the situation described here, shown schematically in Fig.\ref{fig:meanq} (top). The order of the transition however depends on the shape of the function.
First-order transitions can indeed occur if the function $\tilde Q(s)$ is non-convex, in which case one can have multiple equilibrium points, or even first-order transitions.
These first-order transitions are indeed observed experimentally, e.g. in \cite{hochstetteretal}.
 Using the mean field model, this situation is shown in Fig.\ref{fig:meanq} (bottom).

One important issue is when and why the parameter $\chi$ is key to observing these transitions. Let us now extend here, to a more general case, the remarks made in \cite{caravelliscience} about the role of $\chi$. In that case, where we have $\chi=R_{off}-R_{on}/R_{off}$ (analogous to $\chi=(G_{max}-G_{min})/G_{min}$ in this study), 
 the parameter $\chi$ enters in the effective potential multiplying the function $P$. For instance, in the reported experiments of this study, we have $\chi\approx 100$. This implies that in the effective potential of \eqref{eq:landau}, it enters as
\begin{equation}
    Q\Big(P^{-1}\big(\frac{e^s-1}{\chi}\big)\Big)=a v s.
    \label{eq:landau2}
\end{equation}

As a result, the larger the values of $\chi$, the smaller the value of $v$ at which these critical transitions occur. Since typically one has the constraint $\bar r\in [0,1]$, one also must restrict the values of $s\in[s_{max},0]$. This, in turn, restricts the values of $\chi$, which explains why numerically one observes that there exists a minimum value $\chi^*$ in which these transitions occur. In this sense, the amount of memory in the system is an important quantity for these transitions to occur.

\section*{Discussion}

The interplay between nonlinearity, Kirchhoff's laws, and memristive dynamics underlies the observed complex behavior of self-organizing memristive networks. Yet, as we show in this study, because of Kirchhoff's conservation laws much of this complexity can be, at least in the case of two-probe conductance measurements, reabsorbed into the effective parameters of a single junction. This drastic simplification is essentially due to the properties of projector operators.
  
In the present study, we derived an effective mean-field equation describing the  behavior of the effective conductance of a NW network, and the effective equations for a network of nanoparticles.  As we have seen, the dynamical behavior of the effective conductance of a NW network can be well approximated by a mean-field theory, derived from the microscopic equations describing memristive dynamics of a single junction, and the Kirchhoff laws. This is a model that succinctly characterizes the global switching behavior of a memristive NW network. For the case of the experiments presented in this study, it is important to note that the mean-field reduces the system of equations from $4N$ for the case of $N$ junction, to simply $4$, and only $4$ number of free parameters. 
As a result, this study shows that the application of these graph theoretical techniques to a complex system of self-organized NWs provides a quantitative explanation of the response of the system to an applied voltage.

However, most importantly it shows that conductance transitions in NW networks can be explained via the use of effective mean field potentials inspired by the Landau theory of phase transitions. 
This result is the latest addition to a series of papers \cite{caravelliscience,hochstetteretal} showing that there exists a typical phase diagram for the asymptotic conductance or resistance vs applied voltage or current. Whilst the details of the switching region depend on the system under scrutiny, we contend that the seemingly universal properties of these phase diagrams warrant further investigation.
As we have shown analytically and with minimal assumptions, the behavior of the system to the applied voltage can be cast in the form of a standardized mean field equation. The continuous or discontinuous behavior of the conductance of the nanowire connectome is in fact connected to the convexity of the effective potential as a function of the voltage. In particular, we have also provided an analytical explanation for the reason why it is a \textit{generic} feature that these transitions occur in systems with large memory, i.e. when the range of the effective conductance of the system is large. As a result, this study opens a new way of analyzing and classifying the behavior of a generic memristive connectome in terms of the standard theory of phase transitions.

It is worth also mentioning that our graph theoretical techniques have a range of applicability beyond nanowire and nanoparticle networks. For instance, slime molds such as physarum 
polycephalum \cite{Nakagaki2000,Tero_2007,pnasphys}, which inspired a variety of optimization algorithms \cite{adamatzky,BONIFACI2012121}, can indeed be formulated as a memristive component with constraints given by network Kirchhoff's laws induced by the mapping between an incompressible fluid flow and electrical circuits. 

While this work attempts to provide a mean-field theory treatment to memristive devices, it is worth mentioning that the mean-field theory in \cite{caravelliscience} showed the existence of symmetry breaking, while it seems to be absent in our  treatment of memristive nanowires. 
It is also  worth mentioning that our method works within the approximation of discrete memristive junctions, with a voltage drop that can be quantified by a low-dimensional model for the conductance evolution (in our case, one parameter $g$). 
In this framework, it is important to point out that the dynamic behavior of memristive elements composing the network is described by means of a model that, while capturing the main features of dynamics, does not take into account quantum conductance effects that can result in discrete levels of conductance \cite{quantumcond}. Additionally, the model does not take into account disorder due to variability effects in the initial pristine state and in the memristive response of network elements.
Despite these assumptions, the mean-field theory approach is able to describe the main features of the emergent connectome behavior.

In general, while the models we considered in this study are valid for specific initial conditions of  both the nanowire and the nanoparticle states, and within the approximation of homogeneous properties of the single junctions, as discussed  there is a general message that can be evinced from the study of physically relevant connectome models, also based on the discussion of the memristive network toy model introduced in \cite{caravelliscience}.
It is however thus important to stress that more work is needed to bring all these systems into a single theoretical framework. In principle, our techniques could be extended to more complex models such as the one proposed in \cite{ielminisd}, with a continuous family of parameters. Both these extensions will be  the scope of future investigations. 

\newcommand\ackcontent{}

\ifpnas
\section*{Materials and methods}
\showmatmethods{\textit{Sketch of proofs.} 
We derived \eqref{eq:kirchhoff} starting from the equation
\begin{eqnarray}
 \Delta \vec v= {\tilde B}^t(\tilde B\tilde {B}^t+\chi \tilde B \bm{g}\tilde {B}^t)^{-1} \tilde B \vec v_s.
\end{eqnarray}
which can be motivated graph theoretically \cite{caravelliscience}, and then using the identity
\begin{eqnarray}
 {\tilde B}^t(\tilde B\tilde {B}^t+\chi \tilde B  {\bm g}\tilde {B}^t)^{-1} \tilde B=(I+\chi \Omega_{\tilde B} {\bm g})^{-1} \Omega_{\tilde B} \
\end{eqnarray}
from \cite{Caravelli2017}. We used the trick that a conductance in series to the voltage generator (or current generator)

\eqref{eq:vexact} and \eqref{eq:effgeff} were derived by using the block matrix inverse formula for the internal and external voltage drop, and then using the fact that $\Omega^2=\Omega$ in \eqref{eq:sepmat}. For \eqref{eq:effgeff} in particular, we used Ohm's law.

\textit{Fit and numerical integration}. For the fit of eqn. (\ref{fig:fitplotpd}) we integrated numerically \eqref{eq:conddyneff0}-\eqref{eq:conddyneff2} using a Runge-Kutta method of the fourth order as a function of the applied voltage.
} 

\acknow{\ackcontent}
\showacknow{The work of F.C. was conducted under the auspices of the National Nuclear Security Administration
 of the United States Department of Energy at Los Alamos National Laboratory (LANL) under Contract No. DE-AC52-06NA25396 (LANL Laboratory Directed Research Development 20200105ER).} 
 
\section*{Author contributions }{F. Caravelli derived the equations and wrote the initial draft. G. Milano performed the experiments.  All authors contributed equally to the study conception and design, and to the writing of the manuscript.}
\else 
\section*{Acknowledgements}
The work of F.C. was conducted under the auspices of the National Nuclear Security Administration
 of the United States Department of Energy at Los Alamos National Laboratory (LANL) under Contract No. DE-AC52-06NA25396 (LANL Laboratory Directed Research Development 20200105ER)
 \section*{Author contributions }{F. Caravelli derived the equations and wrote the initial draft. G. Milano performed the experiments.  All authors contributed equally to the study conception and design, and to the writing of the manuscript.}
\fi

\bibliography{bibliography}

\begin{thebibliography}{49}%
\makeatletter
\providecommand \@ifxundefined [1]{%
 \@ifx{#1\undefined}
}%
\providecommand \@ifnum [1]{%
 \ifnum #1\expandafter \@firstoftwo
 \else \expandafter \@secondoftwo
 \fi
}%
\providecommand \@ifx [1]{%
 \ifx #1\expandafter \@firstoftwo
 \else \expandafter \@secondoftwo
 \fi
}%
\providecommand \natexlab [1]{#1}%
\providecommand \enquote  [1]{``#1''}%
\providecommand \bibnamefont  [1]{#1}%
\providecommand \bibfnamefont [1]{#1}%
\providecommand \citenamefont [1]{#1}%
\providecommand \href@noop [0]{\@secondoftwo}%
\providecommand \href [0]{\begingroup \@sanitize@url \@href}%
\providecommand \@href[1]{\@@startlink{#1}\@@href}%
\providecommand \@@href[1]{\endgroup#1\@@endlink}%
\providecommand \@sanitize@url [0]{\catcode `\\12\catcode `\$12\catcode
  `\&12\catcode `\#12\catcode `\^12\catcode `\_12\catcode `\%12\relax}%
\providecommand \@@startlink[1]{}%
\providecommand \@@endlink[0]{}%
\providecommand \url  [0]{\begingroup\@sanitize@url \@url }%
\providecommand \@url [1]{\endgroup\@href {#1}{\urlprefix }}%
\providecommand \urlprefix  [0]{URL }%
\providecommand \Eprint [0]{\href }%
\providecommand \doibase [0]{http://dx.doi.org/}%
\providecommand \selectlanguage [0]{\@gobble}%
\providecommand \bibinfo  [0]{\@secondoftwo}%
\providecommand \bibfield  [0]{\@secondoftwo}%
\providecommand \translation [1]{[#1]}%
\providecommand \BibitemOpen [0]{}%
\providecommand \bibitemStop [0]{}%
\providecommand \bibitemNoStop [0]{.\EOS\space}%
\providecommand \EOS [0]{\spacefactor3000\relax}%
\providecommand \BibitemShut  [1]{\csname bibitem#1\endcsname}%
\let\auto@bib@innerbib\@empty
\bibitem [{\citenamefont {Kuncic}\ and\ \citenamefont
  {Nakayama}(2021)}]{Zdenkaadvphys}%
  \BibitemOpen
  \bibfield  {author} {\bibinfo {author} {\bibfnamefont {Z.}~\bibnamefont
  {Kuncic}}\ and\ \bibinfo {author} {\bibfnamefont {T.}~\bibnamefont
  {Nakayama}},\ }\href {\doibase 10.1080/23746149.2021.1894234} {\bibfield
  {journal} {\bibinfo  {journal} {Advances in Physics: X}\ }\textbf {\bibinfo
  {volume} {6}} (\bibinfo {year} {2021}),\
  10.1080/23746149.2021.1894234}\BibitemShut {NoStop}%
\bibitem [{\citenamefont {Jia}\ \emph {et~al.}(2019)\citenamefont {Jia},
  \citenamefont {Lin}, \citenamefont {Huang},\ and\ \citenamefont
  {Duan}}]{elect}%
  \BibitemOpen
  \bibfield  {author} {\bibinfo {author} {\bibfnamefont {C.}~\bibnamefont
  {Jia}}, \bibinfo {author} {\bibfnamefont {Z.}~\bibnamefont {Lin}}, \bibinfo
  {author} {\bibfnamefont {Y.}~\bibnamefont {Huang}}, \ and\ \bibinfo {author}
  {\bibfnamefont {X.}~\bibnamefont {Duan}},\ }\bibfield  {booktitle} {\emph
  {\bibinfo {booktitle} {Chemical Reviews}},\ }\href {\doibase
  10.1021/acs.chemrev.9b00164} {\bibfield  {journal} {\bibinfo  {journal}
  {Chemical Reviews}\ }\textbf {\bibinfo {volume} {119}},\ \bibinfo {pages}
  {9074} (\bibinfo {year} {2019})}\BibitemShut {NoStop}%
\bibitem [{\citenamefont {Yu}\ \emph {et~al.}(2018)\citenamefont {Yu},
  \citenamefont {Pan}, \citenamefont {Zhang}, \citenamefont {Liao},
  \citenamefont {Zhou}, \citenamefont {Yan}, \citenamefont {Xu},\ and\
  \citenamefont {Mai}}]{enstor}%
  \BibitemOpen
  \bibfield  {author} {\bibinfo {author} {\bibfnamefont {K.}~\bibnamefont
  {Yu}}, \bibinfo {author} {\bibfnamefont {X.}~\bibnamefont {Pan}}, \bibinfo
  {author} {\bibfnamefont {G.}~\bibnamefont {Zhang}}, \bibinfo {author}
  {\bibfnamefont {X.}~\bibnamefont {Liao}}, \bibinfo {author} {\bibfnamefont
  {X.}~\bibnamefont {Zhou}}, \bibinfo {author} {\bibfnamefont {M.}~\bibnamefont
  {Yan}}, \bibinfo {author} {\bibfnamefont {L.}~\bibnamefont {Xu}}, \ and\
  \bibinfo {author} {\bibfnamefont {L.}~\bibnamefont {Mai}},\ }\href {\doibase
  https://doi.org/10.1002/aenm.201802369} {\bibfield  {journal} {\bibinfo
  {journal} {Advanced Energy Materials}\ }\textbf {\bibinfo {volume} {8}},\
  \bibinfo {pages} {1802369} (\bibinfo {year} {2018})},\ \Eprint
  {http://arxiv.org/abs/https://onlinelibrary.wiley.com/doi/pdf/10.1002/aenm.201802369}
  {https://onlinelibrary.wiley.com/doi/pdf/10.1002/aenm.201802369} \BibitemShut
  {NoStop}%
\bibitem [{\citenamefont {Patolsky}\ and\ \citenamefont {Lieber}(2005)}]{sens}%
  \BibitemOpen
  \bibfield  {author} {\bibinfo {author} {\bibfnamefont {F.}~\bibnamefont
  {Patolsky}}\ and\ \bibinfo {author} {\bibfnamefont {C.~M.}\ \bibnamefont
  {Lieber}},\ }\href {\doibase https://doi.org/10.1016/S1369-7021(05)00791-1}
  {\bibfield  {journal} {\bibinfo  {journal} {Materials Today}\ }\textbf
  {\bibinfo {volume} {8}},\ \bibinfo {pages} {20} (\bibinfo {year}
  {2005})}\BibitemShut {NoStop}%
\bibitem [{\citenamefont {Milano}\ \emph
  {et~al.}(2022{\natexlab{a}})\citenamefont {Milano}, \citenamefont {Pedretti},
  \citenamefont {Montano}, \citenamefont {Ricci}, \citenamefont {Hashemkhani},
  \citenamefont {Boarino}, \citenamefont {Ielmini},\ and\ \citenamefont
  {Ricciardi}}]{milanores}%
  \BibitemOpen
  \bibfield  {author} {\bibinfo {author} {\bibfnamefont {G.}~\bibnamefont
  {Milano}}, \bibinfo {author} {\bibfnamefont {G.}~\bibnamefont {Pedretti}},
  \bibinfo {author} {\bibfnamefont {K.}~\bibnamefont {Montano}}, \bibinfo
  {author} {\bibfnamefont {S.}~\bibnamefont {Ricci}}, \bibinfo {author}
  {\bibfnamefont {S.}~\bibnamefont {Hashemkhani}}, \bibinfo {author}
  {\bibfnamefont {L.}~\bibnamefont {Boarino}}, \bibinfo {author} {\bibfnamefont
  {D.}~\bibnamefont {Ielmini}}, \ and\ \bibinfo {author} {\bibfnamefont
  {C.}~\bibnamefont {Ricciardi}},\ }\href {\doibase 10.1038/s41563-021-01099-9}
  {\bibfield  {journal} {\bibinfo  {journal} {Nature Materials}\ }\textbf
  {\bibinfo {volume} {21}},\ \bibinfo {pages} {195} (\bibinfo {year}
  {2022}{\natexlab{a}})}\BibitemShut {NoStop}%
\bibitem [{\citenamefont {Diaz-Alvarez}\ \emph {et~al.}(2019)\citenamefont
  {Diaz-Alvarez}, \citenamefont {Higuchi}, \citenamefont {Sanz-Leon},
  \citenamefont {Marcus}, \citenamefont {Shingaya}, \citenamefont {Stieg},
  \citenamefont {Gimzewski}, \citenamefont {Kuncic},\ and\ \citenamefont
  {Nakayama}}]{nakayama}%
  \BibitemOpen
  \bibfield  {author} {\bibinfo {author} {\bibfnamefont {A.}~\bibnamefont
  {Diaz-Alvarez}}, \bibinfo {author} {\bibfnamefont {R.}~\bibnamefont
  {Higuchi}}, \bibinfo {author} {\bibfnamefont {P.}~\bibnamefont {Sanz-Leon}},
  \bibinfo {author} {\bibfnamefont {I.}~\bibnamefont {Marcus}}, \bibinfo
  {author} {\bibfnamefont {Y.}~\bibnamefont {Shingaya}}, \bibinfo {author}
  {\bibfnamefont {A.~Z.}\ \bibnamefont {Stieg}}, \bibinfo {author}
  {\bibfnamefont {J.~K.}\ \bibnamefont {Gimzewski}}, \bibinfo {author}
  {\bibfnamefont {Z.}~\bibnamefont {Kuncic}}, \ and\ \bibinfo {author}
  {\bibfnamefont {T.}~\bibnamefont {Nakayama}},\ }\href {\doibase
  10.1038/s41598-019-51330-6} {\bibfield  {journal} {\bibinfo  {journal}
  {Scientific Reports}\ }\textbf {\bibinfo {volume} {9}},\ \bibinfo {pages}
  {14920} (\bibinfo {year} {2019})}\BibitemShut {NoStop}%
\bibitem [{\citenamefont {Avizienis}\ \emph {et~al.}(2012)\citenamefont
  {Avizienis}, \citenamefont {Sillin}, \citenamefont {Martin-Olmos},
  \citenamefont {Shieh}, \citenamefont {Aono}, \citenamefont {Stieg},\ and\
  \citenamefont {Gimzewski}}]{avizienis}%
  \BibitemOpen
  \bibfield  {author} {\bibinfo {author} {\bibfnamefont {A.}~\bibnamefont
  {Avizienis}}, \bibinfo {author} {\bibfnamefont {H.}~\bibnamefont {Sillin}},
  \bibinfo {author} {\bibfnamefont {C.}~\bibnamefont {Martin-Olmos}}, \bibinfo
  {author} {\bibfnamefont {H.}~\bibnamefont {Shieh}}, \bibinfo {author}
  {\bibfnamefont {M.}~\bibnamefont {Aono}}, \bibinfo {author} {\bibfnamefont
  {A.}~\bibnamefont {Stieg}}, \ and\ \bibinfo {author} {\bibfnamefont
  {J.}~\bibnamefont {Gimzewski}},\ }\href@noop {} {\bibfield  {journal}
  {\bibinfo  {journal} {PLoS One}\ }\textbf {\bibinfo {volume} {7}} (\bibinfo
  {year} {2012})}\BibitemShut {NoStop}%
\bibitem [{\citenamefont {Zhu}\ \emph {et~al.}(2021)\citenamefont {Zhu},
  \citenamefont {Hochstetter}, \citenamefont {Loeffler}, \citenamefont
  {Diaz-Alvarez}, \citenamefont {Nakayama}, \citenamefont {Lizier},\ and\
  \citenamefont {Kuncic}}]{infprop}%
  \BibitemOpen
  \bibfield  {author} {\bibinfo {author} {\bibfnamefont {R.}~\bibnamefont
  {Zhu}}, \bibinfo {author} {\bibfnamefont {J.}~\bibnamefont {Hochstetter}},
  \bibinfo {author} {\bibfnamefont {A.}~\bibnamefont {Loeffler}}, \bibinfo
  {author} {\bibfnamefont {A.}~\bibnamefont {Diaz-Alvarez}}, \bibinfo {author}
  {\bibfnamefont {T.}~\bibnamefont {Nakayama}}, \bibinfo {author}
  {\bibfnamefont {J.~T.}\ \bibnamefont {Lizier}}, \ and\ \bibinfo {author}
  {\bibfnamefont {Z.}~\bibnamefont {Kuncic}},\ }\href {\doibase
  10.1038/s41598-021-92170-7} {\bibfield  {journal} {\bibinfo  {journal}
  {Scientific Reports}\ }\textbf {\bibinfo {volume} {11}},\ \bibinfo {pages}
  {13047} (\bibinfo {year} {2021})}\BibitemShut {NoStop}%
\bibitem [{\citenamefont {Hochstetter}\ \emph {et~al.}(2021)\citenamefont
  {Hochstetter}, \citenamefont {Zhu}, \citenamefont {Loeffler}, \citenamefont
  {Diaz-Alvarez}, \citenamefont {Nakayama},\ and\ \citenamefont
  {Kuncic}}]{hochstetteretal}%
  \BibitemOpen
  \bibfield  {author} {\bibinfo {author} {\bibfnamefont {J.}~\bibnamefont
  {Hochstetter}}, \bibinfo {author} {\bibfnamefont {R.}~\bibnamefont {Zhu}},
  \bibinfo {author} {\bibfnamefont {A.}~\bibnamefont {Loeffler}}, \bibinfo
  {author} {\bibfnamefont {A.}~\bibnamefont {Diaz-Alvarez}}, \bibinfo {author}
  {\bibfnamefont {T.}~\bibnamefont {Nakayama}}, \ and\ \bibinfo {author}
  {\bibfnamefont {Z.}~\bibnamefont {Kuncic}},\ }\href {\doibase
  10.1038/s41467-021-24260-z} {\bibfield  {journal} {\bibinfo  {journal}
  {Nature Communications}\ }\textbf {\bibinfo {volume} {12}},\ \bibinfo {pages}
  {4008} (\bibinfo {year} {2021})}\BibitemShut {NoStop}%
\bibitem [{\citenamefont {Milano}\ \emph
  {et~al.}(2022{\natexlab{b}})\citenamefont {Milano}, \citenamefont {Miranda},\
  and\ \citenamefont {Ricciardi}}]{MILANO2022137}%
  \BibitemOpen
  \bibfield  {author} {\bibinfo {author} {\bibfnamefont {G.}~\bibnamefont
  {Milano}}, \bibinfo {author} {\bibfnamefont {E.}~\bibnamefont {Miranda}}, \
  and\ \bibinfo {author} {\bibfnamefont {C.}~\bibnamefont {Ricciardi}},\ }\href
  {\doibase https://doi.org/10.1016/j.neunet.2022.02.022} {\bibfield  {journal}
  {\bibinfo  {journal} {Neural Networks}\ }\textbf {\bibinfo {volume} {150}},\
  \bibinfo {pages} {137} (\bibinfo {year} {2022}{\natexlab{b}})}\BibitemShut
  {NoStop}%
\bibitem [{\citenamefont {Milano}\ \emph {et~al.}(2019)\citenamefont {Milano},
  \citenamefont {Porro}, \citenamefont {Valov},\ and\ \citenamefont
  {Ricciardi}}]{Milano2019}%
  \BibitemOpen
  \bibfield  {author} {\bibinfo {author} {\bibfnamefont {G.}~\bibnamefont
  {Milano}}, \bibinfo {author} {\bibfnamefont {S.}~\bibnamefont {Porro}},
  \bibinfo {author} {\bibfnamefont {I.}~\bibnamefont {Valov}}, \ and\ \bibinfo
  {author} {\bibfnamefont {C.}~\bibnamefont {Ricciardi}},\ }\href {\doibase
  10.1002/aelm.201800909} {\bibfield  {journal} {\bibinfo  {journal} {Advanced
  Electronic Materials}\ }\textbf {\bibinfo {volume} {5}},\ \bibinfo {pages}
  {1800909} (\bibinfo {year} {2019})}\BibitemShut {NoStop}%
\bibitem [{\citenamefont {Manning}\ \emph {et~al.}(2018)\citenamefont
  {Manning}, \citenamefont {Niosi}, \citenamefont {da~Rocha}, \citenamefont
  {Bellew}, \citenamefont {O’Callaghan}, \citenamefont {Biswas},
  \citenamefont {Flowers}, \citenamefont {Wiley}, \citenamefont {Holmes},
  \citenamefont {Ferreira},\ and\ \citenamefont {Boland}}]{Manning2018}%
  \BibitemOpen
  \bibfield  {author} {\bibinfo {author} {\bibfnamefont {H.~G.}\ \bibnamefont
  {Manning}}, \bibinfo {author} {\bibfnamefont {F.}~\bibnamefont {Niosi}},
  \bibinfo {author} {\bibfnamefont {C.~G.}\ \bibnamefont {da~Rocha}}, \bibinfo
  {author} {\bibfnamefont {A.~T.}\ \bibnamefont {Bellew}}, \bibinfo {author}
  {\bibfnamefont {C.}~\bibnamefont {O’Callaghan}}, \bibinfo {author}
  {\bibfnamefont {S.}~\bibnamefont {Biswas}}, \bibinfo {author} {\bibfnamefont
  {P.~F.}\ \bibnamefont {Flowers}}, \bibinfo {author} {\bibfnamefont {B.~J.}\
  \bibnamefont {Wiley}}, \bibinfo {author} {\bibfnamefont {J.~D.}\ \bibnamefont
  {Holmes}}, \bibinfo {author} {\bibfnamefont {M.~S.}\ \bibnamefont
  {Ferreira}}, \ and\ \bibinfo {author} {\bibfnamefont {J.~J.}\ \bibnamefont
  {Boland}},\ }\href {\doibase 10.1038/s41467-018-05517-6} {\bibfield
  {journal} {\bibinfo  {journal} {Nature Communications}\ }\textbf {\bibinfo
  {volume} {9}},\ \bibinfo {pages} {3219} (\bibinfo {year} {2018})}\BibitemShut
  {NoStop}%
\bibitem [{\citenamefont {Milano}\ \emph
  {et~al.}(2020{\natexlab{a}})\citenamefont {Milano}, \citenamefont {Pedretti},
  \citenamefont {Fretto}, \citenamefont {Boarino}, \citenamefont {Benfenati},
  \citenamefont {Ielmini}, \citenamefont {Valov},\ and\ \citenamefont
  {Ricciardi}}]{milano2020}%
  \BibitemOpen
  \bibfield  {author} {\bibinfo {author} {\bibfnamefont {G.}~\bibnamefont
  {Milano}}, \bibinfo {author} {\bibfnamefont {G.}~\bibnamefont {Pedretti}},
  \bibinfo {author} {\bibfnamefont {M.}~\bibnamefont {Fretto}}, \bibinfo
  {author} {\bibfnamefont {L.}~\bibnamefont {Boarino}}, \bibinfo {author}
  {\bibfnamefont {F.}~\bibnamefont {Benfenati}}, \bibinfo {author}
  {\bibfnamefont {D.}~\bibnamefont {Ielmini}}, \bibinfo {author} {\bibfnamefont
  {I.}~\bibnamefont {Valov}}, \ and\ \bibinfo {author} {\bibfnamefont
  {C.}~\bibnamefont {Ricciardi}},\ }\href {\doibase
  https://doi.org/10.1002/aisy.202000096} {\bibfield  {journal} {\bibinfo
  {journal} {Advanced Intelligent Systems}\ }\textbf {\bibinfo {volume} {2}},\
  \bibinfo {pages} {2000096} (\bibinfo {year}
  {2020}{\natexlab{a}})}\BibitemShut {NoStop}%
\bibitem [{\citenamefont {Nagashima}\ \emph {et~al.}(2011)\citenamefont
  {Nagashima}, \citenamefont {Yanagida}, \citenamefont {Oka}, \citenamefont
  {Kanai}, \citenamefont {Klamchuen}, \citenamefont {Kim}, \citenamefont
  {Park},\ and\ \citenamefont {Kawai}}]{memr1}%
  \BibitemOpen
  \bibfield  {author} {\bibinfo {author} {\bibfnamefont {K.}~\bibnamefont
  {Nagashima}}, \bibinfo {author} {\bibfnamefont {T.}~\bibnamefont {Yanagida}},
  \bibinfo {author} {\bibfnamefont {K.}~\bibnamefont {Oka}}, \bibinfo {author}
  {\bibfnamefont {M.}~\bibnamefont {Kanai}}, \bibinfo {author} {\bibfnamefont
  {A.}~\bibnamefont {Klamchuen}}, \bibinfo {author} {\bibfnamefont {J.-S.}\
  \bibnamefont {Kim}}, \bibinfo {author} {\bibfnamefont {B.~H.}\ \bibnamefont
  {Park}}, \ and\ \bibinfo {author} {\bibfnamefont {T.}~\bibnamefont {Kawai}},\
  }\bibfield  {booktitle} {\emph {\bibinfo {booktitle} {Nano Letters}},\ }\href
  {\doibase 10.1021/nl200707n} {\bibfield  {journal} {\bibinfo  {journal} {Nano
  Letters}\ }\textbf {\bibinfo {volume} {11}},\ \bibinfo {pages} {2114}
  (\bibinfo {year} {2011})}\BibitemShut {NoStop}%
\bibitem [{\citenamefont {He}\ \emph {et~al.}(2011)\citenamefont {He},
  \citenamefont {Liao}, \citenamefont {Wu}, \citenamefont {Tian}, \citenamefont
  {Xu}, \citenamefont {Cross}, \citenamefont {Duesberg}, \citenamefont
  {Shvets},\ and\ \citenamefont {Yu}}]{memr2}%
  \BibitemOpen
  \bibfield  {author} {\bibinfo {author} {\bibfnamefont {L.}~\bibnamefont
  {He}}, \bibinfo {author} {\bibfnamefont {Z.-M.}\ \bibnamefont {Liao}},
  \bibinfo {author} {\bibfnamefont {H.-C.}\ \bibnamefont {Wu}}, \bibinfo
  {author} {\bibfnamefont {X.-X.}\ \bibnamefont {Tian}}, \bibinfo {author}
  {\bibfnamefont {D.-S.}\ \bibnamefont {Xu}}, \bibinfo {author} {\bibfnamefont
  {G.~L.~W.}\ \bibnamefont {Cross}}, \bibinfo {author} {\bibfnamefont {G.~S.}\
  \bibnamefont {Duesberg}}, \bibinfo {author} {\bibfnamefont {I.~V.}\
  \bibnamefont {Shvets}}, \ and\ \bibinfo {author} {\bibfnamefont {D.-P.}\
  \bibnamefont {Yu}},\ }\bibfield  {booktitle} {\emph {\bibinfo {booktitle}
  {Nano Letters}},\ }\href {\doibase 10.1021/nl202017k} {\bibfield  {journal}
  {\bibinfo  {journal} {Nano Letters}\ }\textbf {\bibinfo {volume} {11}},\
  \bibinfo {pages} {4601} (\bibinfo {year} {2011})}\BibitemShut {NoStop}%
\bibitem [{\citenamefont {Yang}\ \emph {et~al.}(2011)\citenamefont {Yang},
  \citenamefont {Zhang}, \citenamefont {Gao}, \citenamefont {Zeng},
  \citenamefont {Zhou}, \citenamefont {Xie},\ and\ \citenamefont {Pan}}]{yang}%
  \BibitemOpen
  \bibfield  {author} {\bibinfo {author} {\bibfnamefont {Y.}~\bibnamefont
  {Yang}}, \bibinfo {author} {\bibfnamefont {X.}~\bibnamefont {Zhang}},
  \bibinfo {author} {\bibfnamefont {M.}~\bibnamefont {Gao}}, \bibinfo {author}
  {\bibfnamefont {F.}~\bibnamefont {Zeng}}, \bibinfo {author} {\bibfnamefont
  {W.}~\bibnamefont {Zhou}}, \bibinfo {author} {\bibfnamefont {S.}~\bibnamefont
  {Xie}}, \ and\ \bibinfo {author} {\bibfnamefont {F.}~\bibnamefont {Pan}},\
  }\href {\doibase 10.1039/C1NR10096C} {\bibfield  {journal} {\bibinfo
  {journal} {Nanoscale}\ }\textbf {\bibinfo {volume} {3}},\ \bibinfo {pages}
  {1917} (\bibinfo {year} {2011})}\BibitemShut {NoStop}%
\bibitem [{\citenamefont {Mead}(1990)}]{mead}%
  \BibitemOpen
  \bibfield  {author} {\bibinfo {author} {\bibfnamefont {C.}~\bibnamefont
  {Mead}},\ }\href {\doibase 10.1109/5.58356} {\bibfield  {journal} {\bibinfo
  {journal} {Proceedings of the IEEE}\ }\textbf {\bibinfo {volume} {78}},\
  \bibinfo {pages} {1629} (\bibinfo {year} {1990})}\BibitemShut {NoStop}%
\bibitem [{\citenamefont {S.}\ \emph {et~al.}(2008)\citenamefont {S.},
  \citenamefont {S.}, \citenamefont {S.},\ and\ \citenamefont
  {Williams}}]{stru08}%
  \BibitemOpen
  \bibfield  {author} {\bibinfo {author} {\bibfnamefont {D.~B.}\ \bibnamefont
  {S.}}, \bibinfo {author} {\bibfnamefont {G.~S.}\ \bibnamefont {S.}}, \bibinfo
  {author} {\bibfnamefont {D.~R.}\ \bibnamefont {S.}}, \ and\ \bibinfo {author}
  {\bibfnamefont {R.~S.}\ \bibnamefont {Williams}},\ }\href {\doibase
  10.1038/nature06932} {\bibfield  {journal} {\bibinfo  {journal} {Nature}\
  }\textbf {\bibinfo {volume} {453}},\ \bibinfo {pages} {80} (\bibinfo {year}
  {2008})}\BibitemShut {NoStop}%
\bibitem [{\citenamefont {Zhang}\ \emph {et~al.}(2020)\citenamefont {Zhang},
  \citenamefont {Wang}, \citenamefont {Zhu}, \citenamefont {Yang},
  \citenamefont {Rao}, \citenamefont {Song}, \citenamefont {Zhuo},
  \citenamefont {Zhang}, \citenamefont {Cui}, \citenamefont {Shen},
  \citenamefont {Huang},\ and\ \citenamefont {Yang}}]{Zhang2020}%
  \BibitemOpen
  \bibfield  {author} {\bibinfo {author} {\bibfnamefont {Y.}~\bibnamefont
  {Zhang}}, \bibinfo {author} {\bibfnamefont {Z.}~\bibnamefont {Wang}},
  \bibinfo {author} {\bibfnamefont {J.}~\bibnamefont {Zhu}}, \bibinfo {author}
  {\bibfnamefont {Y.}~\bibnamefont {Yang}}, \bibinfo {author} {\bibfnamefont
  {M.}~\bibnamefont {Rao}}, \bibinfo {author} {\bibfnamefont {W.}~\bibnamefont
  {Song}}, \bibinfo {author} {\bibfnamefont {Y.}~\bibnamefont {Zhuo}}, \bibinfo
  {author} {\bibfnamefont {X.}~\bibnamefont {Zhang}}, \bibinfo {author}
  {\bibfnamefont {M.}~\bibnamefont {Cui}}, \bibinfo {author} {\bibfnamefont
  {L.}~\bibnamefont {Shen}}, \bibinfo {author} {\bibfnamefont {R.}~\bibnamefont
  {Huang}}, \ and\ \bibinfo {author} {\bibfnamefont {J.~J.}\ \bibnamefont
  {Yang}},\ }\href {\doibase 10.1063/1.5124027} {\bibfield  {journal} {\bibinfo
   {journal} {Applied Physics Reviews}\ }\textbf {\bibinfo {volume} {7}},\
  \bibinfo {pages} {011308} (\bibinfo {year} {2020})}\BibitemShut {NoStop}%
\bibitem [{\citenamefont {Xia}\ and\ \citenamefont {Yang}(2019)}]{Xia2019}%
  \BibitemOpen
  \bibfield  {author} {\bibinfo {author} {\bibfnamefont {Q.}~\bibnamefont
  {Xia}}\ and\ \bibinfo {author} {\bibfnamefont {J.~J.}\ \bibnamefont {Yang}},\
  }\href {\doibase 10.1038/s41563-019-0291-x} {\bibfield  {journal} {\bibinfo
  {journal} {Nature Materials}\ }\textbf {\bibinfo {volume} {18}},\ \bibinfo
  {pages} {309} (\bibinfo {year} {2019})}\BibitemShut {NoStop}%
\bibitem [{\citenamefont {Caravelli}\ and\ \citenamefont
  {Carbajal}(2018)}]{reviewCarCar}%
  \BibitemOpen
  \bibfield  {author} {\bibinfo {author} {\bibfnamefont {F.}~\bibnamefont
  {Caravelli}}\ and\ \bibinfo {author} {\bibfnamefont {J.}~\bibnamefont
  {Carbajal}},\ }\href {\doibase 10.3390/technologies6040118} {\bibfield
  {journal} {\bibinfo  {journal} {Technologies}\ }\textbf {\bibinfo {volume}
  {6}},\ \bibinfo {pages} {118} (\bibinfo {year} {2018})}\BibitemShut {NoStop}%
\bibitem [{\citenamefont {Saccone}\ \emph {et~al.}(2022)\citenamefont
  {Saccone}, \citenamefont {Caravelli}, \citenamefont {Hofhuis}, \citenamefont
  {Parchenko}, \citenamefont {Birkh{\"o}lzer}, \citenamefont {Dhuey},
  \citenamefont {Kleibert}, \citenamefont {van Dijken}, \citenamefont
  {Nisoli},\ and\ \citenamefont {Farhan}}]{saccone1}%
  \BibitemOpen
  \bibfield  {author} {\bibinfo {author} {\bibfnamefont {M.}~\bibnamefont
  {Saccone}}, \bibinfo {author} {\bibfnamefont {F.}~\bibnamefont {Caravelli}},
  \bibinfo {author} {\bibfnamefont {K.}~\bibnamefont {Hofhuis}}, \bibinfo
  {author} {\bibfnamefont {S.}~\bibnamefont {Parchenko}}, \bibinfo {author}
  {\bibfnamefont {Y.~A.}\ \bibnamefont {Birkh{\"o}lzer}}, \bibinfo {author}
  {\bibfnamefont {S.}~\bibnamefont {Dhuey}}, \bibinfo {author} {\bibfnamefont
  {A.}~\bibnamefont {Kleibert}}, \bibinfo {author} {\bibfnamefont
  {S.}~\bibnamefont {van Dijken}}, \bibinfo {author} {\bibfnamefont
  {C.}~\bibnamefont {Nisoli}}, \ and\ \bibinfo {author} {\bibfnamefont
  {A.}~\bibnamefont {Farhan}},\ }\href {\doibase 10.1038/s41567-022-01538-7}
  {\bibfield  {journal} {\bibinfo  {journal} {Nature Physics}\ }\textbf
  {\bibinfo {volume} {18}},\ \bibinfo {pages} {517} (\bibinfo {year}
  {2022})}\BibitemShut {NoStop}%
\bibitem [{\citenamefont {Caravelli}\ \emph {et~al.}(2022)\citenamefont
  {Caravelli}, \citenamefont {Chern},\ and\ \citenamefont
  {Nisoli}}]{caravelli_2022}%
  \BibitemOpen
  \bibfield  {author} {\bibinfo {author} {\bibfnamefont {F.}~\bibnamefont
  {Caravelli}}, \bibinfo {author} {\bibfnamefont {G.-W.}\ \bibnamefont
  {Chern}}, \ and\ \bibinfo {author} {\bibfnamefont {C.}~\bibnamefont
  {Nisoli}},\ }\href {\doibase 10.1088/1367-2630/ac4c0a} {\bibfield  {journal}
  {\bibinfo  {journal} {New Journal of Physics}\ }\textbf {\bibinfo {volume}
  {24}},\ \bibinfo {pages} {023020} (\bibinfo {year} {2022})}\BibitemShut
  {NoStop}%
\bibitem [{\citenamefont {Caravelli}\ and\ \citenamefont
  {Nisoli}(2020)}]{caravelli_2020}%
  \BibitemOpen
  \bibfield  {author} {\bibinfo {author} {\bibfnamefont {F.}~\bibnamefont
  {Caravelli}}\ and\ \bibinfo {author} {\bibfnamefont {C.}~\bibnamefont
  {Nisoli}},\ }\href {\doibase 10.1088/1367-2630/abbf21} {\bibfield  {journal}
  {\bibinfo  {journal} {New Journal of Physics}\ }\textbf {\bibinfo {volume}
  {22}},\ \bibinfo {pages} {103052} (\bibinfo {year} {2020})}\BibitemShut
  {NoStop}%
\bibitem [{\citenamefont {Gartside}\ \emph {et~al.}(2022)\citenamefont
  {Gartside}, \citenamefont {Stenning}, \citenamefont {Vanstone}, \citenamefont
  {Holder}, \citenamefont {Arroo}, \citenamefont {Dion}, \citenamefont
  {Caravelli}, \citenamefont {Kurebayashi},\ and\ \citenamefont
  {Branford}}]{gartside}%
  \BibitemOpen
  \bibfield  {author} {\bibinfo {author} {\bibfnamefont {J.~C.}\ \bibnamefont
  {Gartside}}, \bibinfo {author} {\bibfnamefont {K.~D.}\ \bibnamefont
  {Stenning}}, \bibinfo {author} {\bibfnamefont {A.}~\bibnamefont {Vanstone}},
  \bibinfo {author} {\bibfnamefont {H.~H.}\ \bibnamefont {Holder}}, \bibinfo
  {author} {\bibfnamefont {D.~M.}\ \bibnamefont {Arroo}}, \bibinfo {author}
  {\bibfnamefont {T.}~\bibnamefont {Dion}}, \bibinfo {author} {\bibfnamefont
  {F.}~\bibnamefont {Caravelli}}, \bibinfo {author} {\bibfnamefont
  {H.}~\bibnamefont {Kurebayashi}}, \ and\ \bibinfo {author} {\bibfnamefont
  {W.~R.}\ \bibnamefont {Branford}},\ }\href {\doibase
  10.1038/s41565-022-01091-7} {\bibfield  {journal} {\bibinfo  {journal}
  {Nature Nanotechnology}\ }\textbf {\bibinfo {volume} {17}},\ \bibinfo {pages}
  {460} (\bibinfo {year} {2022})}\BibitemShut {NoStop}%
\bibitem [{\citenamefont {Zucker}\ and\ \citenamefont {Regehr}(2002)}]{plast}%
  \BibitemOpen
  \bibfield  {author} {\bibinfo {author} {\bibfnamefont {R.~S.}\ \bibnamefont
  {Zucker}}\ and\ \bibinfo {author} {\bibfnamefont {W.~G.}\ \bibnamefont
  {Regehr}},\ }\href {\doibase 10.1146/annurev.physiol.64.092501.114547}
  {\bibfield  {journal} {\bibinfo  {journal} {Annual Review of Physiology}\
  }\textbf {\bibinfo {volume} {64}},\ \bibinfo {pages} {355} (\bibinfo {year}
  {2002})},\ \bibinfo {note} {pMID: 11826273},\ \Eprint
  {http://arxiv.org/abs/https://doi.org/10.1146/annurev.physiol.64.092501.114547}
  {https://doi.org/10.1146/annurev.physiol.64.092501.114547} \BibitemShut
  {NoStop}%
\bibitem [{\citenamefont {Ohno}\ \emph {et~al.}(2011)\citenamefont {Ohno},
  \citenamefont {Hasegawa}, \citenamefont {Nayak}, \citenamefont {Tsuruoka},
  \citenamefont {Gimzewski},\ and\ \citenamefont {Aono}}]{AtomicSwitch1}%
  \BibitemOpen
  \bibfield  {author} {\bibinfo {author} {\bibfnamefont {T.}~\bibnamefont
  {Ohno}}, \bibinfo {author} {\bibfnamefont {T.}~\bibnamefont {Hasegawa}},
  \bibinfo {author} {\bibfnamefont {A.}~\bibnamefont {Nayak}}, \bibinfo
  {author} {\bibfnamefont {T.}~\bibnamefont {Tsuruoka}}, \bibinfo {author}
  {\bibfnamefont {J.~K.}\ \bibnamefont {Gimzewski}}, \ and\ \bibinfo {author}
  {\bibfnamefont {M.}~\bibnamefont {Aono}},\ }\href {\doibase
  10.1063/1.3662390} {\bibfield  {journal} {\bibinfo  {journal} {Applied
  Physics Letters}\ }\textbf {\bibinfo {volume} {99}},\ \bibinfo {pages}
  {203108} (\bibinfo {year} {2011})}\BibitemShut {NoStop}%
\bibitem [{\citenamefont {Wang}\ \emph {et~al.}(2016)\citenamefont {Wang},
  \citenamefont {Joshi}, \citenamefont {Savel'ev}, \citenamefont {Jiang},
  \citenamefont {Midya}, \citenamefont {Lin}, \citenamefont {Hu}, \citenamefont
  {Ge}, \citenamefont {Strachan}, \citenamefont {Li}, \citenamefont {Wu},
  \citenamefont {Barnell}, \citenamefont {Li}, \citenamefont {Xin},
  \citenamefont {Williams}, \citenamefont {Xia},\ and\ \citenamefont
  {Yang}}]{AtomicSwitch2}%
  \BibitemOpen
  \bibfield  {author} {\bibinfo {author} {\bibfnamefont {Z.}~\bibnamefont
  {Wang}}, \bibinfo {author} {\bibfnamefont {S.}~\bibnamefont {Joshi}},
  \bibinfo {author} {\bibfnamefont {S.~E.}\ \bibnamefont {Savel'ev}}, \bibinfo
  {author} {\bibfnamefont {H.}~\bibnamefont {Jiang}}, \bibinfo {author}
  {\bibfnamefont {R.}~\bibnamefont {Midya}}, \bibinfo {author} {\bibfnamefont
  {P.}~\bibnamefont {Lin}}, \bibinfo {author} {\bibfnamefont {M.}~\bibnamefont
  {Hu}}, \bibinfo {author} {\bibfnamefont {N.}~\bibnamefont {Ge}}, \bibinfo
  {author} {\bibfnamefont {J.~P.}\ \bibnamefont {Strachan}}, \bibinfo {author}
  {\bibfnamefont {Z.}~\bibnamefont {Li}}, \bibinfo {author} {\bibfnamefont
  {Q.}~\bibnamefont {Wu}}, \bibinfo {author} {\bibfnamefont {M.}~\bibnamefont
  {Barnell}}, \bibinfo {author} {\bibfnamefont {G.-L.}\ \bibnamefont {Li}},
  \bibinfo {author} {\bibfnamefont {H.~L.}\ \bibnamefont {Xin}}, \bibinfo
  {author} {\bibfnamefont {R.~S.}\ \bibnamefont {Williams}}, \bibinfo {author}
  {\bibfnamefont {Q.}~\bibnamefont {Xia}}, \ and\ \bibinfo {author}
  {\bibfnamefont {J.~J.}\ \bibnamefont {Yang}},\ }\href {\doibase
  10.1038/nmat4756} {\bibfield  {journal} {\bibinfo  {journal} {Nature
  Materials}\ }\textbf {\bibinfo {volume} {16}},\ \bibinfo {pages} {101}
  (\bibinfo {year} {2016})}\BibitemShut {NoStop}%
\bibitem [{\citenamefont {Milano}\ \emph {et~al.}(2018)\citenamefont {Milano},
  \citenamefont {Luebben}, \citenamefont {Ma}, \citenamefont {Dunin-Borkowski},
  \citenamefont {Boarino}, \citenamefont {Pirri}, \citenamefont {Waser},
  \citenamefont {Ricciardi},\ and\ \citenamefont {Valov}}]{milanor}%
  \BibitemOpen
  \bibfield  {author} {\bibinfo {author} {\bibfnamefont {G.}~\bibnamefont
  {Milano}}, \bibinfo {author} {\bibfnamefont {M.}~\bibnamefont {Luebben}},
  \bibinfo {author} {\bibfnamefont {Z.}~\bibnamefont {Ma}}, \bibinfo {author}
  {\bibfnamefont {R.}~\bibnamefont {Dunin-Borkowski}}, \bibinfo {author}
  {\bibfnamefont {L.}~\bibnamefont {Boarino}}, \bibinfo {author} {\bibfnamefont
  {C.~F.}\ \bibnamefont {Pirri}}, \bibinfo {author} {\bibfnamefont
  {R.}~\bibnamefont {Waser}}, \bibinfo {author} {\bibfnamefont
  {C.}~\bibnamefont {Ricciardi}}, \ and\ \bibinfo {author} {\bibfnamefont
  {I.}~\bibnamefont {Valov}},\ }\href {\doibase 10.1038/s41467-018-07330-7}
  {\bibfield  {journal} {\bibinfo  {journal} {Nature Communications}\ }\textbf
  {\bibinfo {volume} {9}},\ \bibinfo {pages} {5151} (\bibinfo {year}
  {2018})}\BibitemShut {NoStop}%
\bibitem [{\citenamefont {Sheldon}\ \emph {et~al.}(2022)\citenamefont
  {Sheldon}, \citenamefont {Kolchinsky},\ and\ \citenamefont
  {Caravelli}}]{sheldon}%
  \BibitemOpen
  \bibfield  {author} {\bibinfo {author} {\bibfnamefont {F.~C.}\ \bibnamefont
  {Sheldon}}, \bibinfo {author} {\bibfnamefont {A.}~\bibnamefont {Kolchinsky}},
  \ and\ \bibinfo {author} {\bibfnamefont {F.}~\bibnamefont {Caravelli}},\
  }\href {\doibase 10.1103/PhysRevE.106.045310} {\bibfield  {journal} {\bibinfo
   {journal} {Phys. Rev. E}\ }\textbf {\bibinfo {volume} {106}},\ \bibinfo
  {pages} {045310} (\bibinfo {year} {2022})}\BibitemShut {NoStop}%
\bibitem [{\citenamefont {Loeffler}\ \emph {et~al.}(2020)\citenamefont
  {Loeffler}, \citenamefont {Zhu}, \citenamefont {Hochstetter}, \citenamefont
  {Li}, \citenamefont {Fu}, \citenamefont {Diaz-Alvarez}, \citenamefont
  {Nakayama}, \citenamefont {Shine},\ and\ \citenamefont
  {Kuncic}}]{loeffler2020}%
  \BibitemOpen
  \bibfield  {author} {\bibinfo {author} {\bibfnamefont {A.}~\bibnamefont
  {Loeffler}}, \bibinfo {author} {\bibfnamefont {R.}~\bibnamefont {Zhu}},
  \bibinfo {author} {\bibfnamefont {J.}~\bibnamefont {Hochstetter}}, \bibinfo
  {author} {\bibfnamefont {M.}~\bibnamefont {Li}}, \bibinfo {author}
  {\bibfnamefont {K.}~\bibnamefont {Fu}}, \bibinfo {author} {\bibfnamefont
  {A.}~\bibnamefont {Diaz-Alvarez}}, \bibinfo {author} {\bibfnamefont
  {T.}~\bibnamefont {Nakayama}}, \bibinfo {author} {\bibfnamefont {J.~M.}\
  \bibnamefont {Shine}}, \ and\ \bibinfo {author} {\bibfnamefont
  {Z.}~\bibnamefont {Kuncic}},\ }\href {\doibase 10.3389/fnins.2020.00184}
  {\bibfield  {journal} {\bibinfo  {journal} {Frontiers in Neuroscience}\
  }\textbf {\bibinfo {volume} {14}} (\bibinfo {year} {2020}),\
  10.3389/fnins.2020.00184}\BibitemShut {NoStop}%
\bibitem [{\citenamefont {Caravelli}\ \emph {et~al.}(2021)\citenamefont
  {Caravelli}, \citenamefont {Sheldon},\ and\ \citenamefont
  {Traversa}}]{caravelliscience}%
  \BibitemOpen
  \bibfield  {author} {\bibinfo {author} {\bibfnamefont {F.}~\bibnamefont
  {Caravelli}}, \bibinfo {author} {\bibfnamefont {F.~C.}\ \bibnamefont
  {Sheldon}}, \ and\ \bibinfo {author} {\bibfnamefont {F.~L.}\ \bibnamefont
  {Traversa}},\ }\href {\doibase 10.1126/sciadv.abh1542} {\bibfield  {journal}
  {\bibinfo  {journal} {Science Advances}\ }\textbf {\bibinfo {volume} {7}}
  (\bibinfo {year} {2021}),\ 10.1126/sciadv.abh1542}\BibitemShut {NoStop}%
\bibitem [{\citenamefont {Yang}\ \emph {et~al.}(2012)\citenamefont {Yang},
  \citenamefont {Strukov},\ and\ \citenamefont {Stewart}}]{stru}%
  \BibitemOpen
  \bibfield  {author} {\bibinfo {author} {\bibfnamefont {J.~J.}\ \bibnamefont
  {Yang}}, \bibinfo {author} {\bibfnamefont {D.~B.}\ \bibnamefont {Strukov}}, \
  and\ \bibinfo {author} {\bibfnamefont {D.~R.}\ \bibnamefont {Stewart}},\
  }\href {\doibase 10.1038/nnano.2012.240} {\bibfield  {journal} {\bibinfo
  {journal} {Nature Nanotechnology}\ }\textbf {\bibinfo {volume} {8}},\
  \bibinfo {pages} {13} (\bibinfo {year} {2012})}\BibitemShut {NoStop}%
\bibitem [{\citenamefont {Miranda}\ \emph {et~al.}(2020)\citenamefont
  {Miranda}, \citenamefont {Milano},\ and\ \citenamefont
  {Ricciardi}}]{Mirandeetal}%
  \BibitemOpen
  \bibfield  {author} {\bibinfo {author} {\bibfnamefont {E.}~\bibnamefont
  {Miranda}}, \bibinfo {author} {\bibfnamefont {G.}~\bibnamefont {Milano}}, \
  and\ \bibinfo {author} {\bibfnamefont {C.}~\bibnamefont {Ricciardi}},\ }\href
  {\doibase 10.1109/TNANO.2020.3009734} {\bibfield  {journal} {\bibinfo
  {journal} {IEEE Transactions on Nanotechnology}\ }\textbf {\bibinfo {volume}
  {19}} (\bibinfo {year} {2020}),\ 10.1109/TNANO.2020.3009734}\BibitemShut
  {NoStop}%
\bibitem [{\citenamefont {Zegarac}\ and\ \citenamefont
  {Caravelli}(2019)}]{Caravelli2019}%
  \BibitemOpen
  \bibfield  {author} {\bibinfo {author} {\bibfnamefont {A.}~\bibnamefont
  {Zegarac}}\ and\ \bibinfo {author} {\bibfnamefont {F.}~\bibnamefont
  {Caravelli}},\ }\href {\doibase 10.1209/0295-5075/125/10001} {\bibfield
  {journal} {\bibinfo  {journal} {{EPL} (Europhysics Letters)}\ }\textbf
  {\bibinfo {volume} {125}},\ \bibinfo {pages} {10001} (\bibinfo {year}
  {2019})}\BibitemShut {NoStop}%
\bibitem [{\citenamefont {Caravelli}\ \emph {et~al.}(2017)\citenamefont
  {Caravelli}, \citenamefont {Traversa},\ and\ \citenamefont {{Di
  Ventra}}}]{caravelli2016rl}%
  \BibitemOpen
  \bibfield  {author} {\bibinfo {author} {\bibfnamefont {F.}~\bibnamefont
  {Caravelli}}, \bibinfo {author} {\bibfnamefont {F.~L.}\ \bibnamefont
  {Traversa}}, \ and\ \bibinfo {author} {\bibfnamefont {M.}~\bibnamefont {{Di
  Ventra}}},\ }\href {\doibase 10.1103/physreve.95.022140} {\bibfield
  {journal} {\bibinfo  {journal} {Physical Review E}\ }\textbf {\bibinfo
  {volume} {95}},\ \bibinfo {pages} {022140} (\bibinfo {year}
  {2017})}\BibitemShut {NoStop}%
\bibitem [{\citenamefont {Caravelli}(2017)}]{Caravelli2017}%
  \BibitemOpen
  \bibfield  {author} {\bibinfo {author} {\bibfnamefont {F.}~\bibnamefont
  {Caravelli}},\ }\href {\doibase 10.1103/physreve.96.052206} {\bibfield
  {journal} {\bibinfo  {journal} {Physical Review E}\ }\textbf {\bibinfo
  {volume} {96}},\ \bibinfo {pages} {052206} (\bibinfo {year}
  {2017})}\BibitemShut {NoStop}%
\bibitem [{\citenamefont {Milano}\ \emph
  {et~al.}(2020{\natexlab{b}})\citenamefont {Milano}, \citenamefont {Pedretti},
  \citenamefont {Fretto}, \citenamefont {Boarino}, \citenamefont {Benfenati},
  \citenamefont {Ielmini}, \citenamefont {Valov},\ and\ \citenamefont
  {Ricciardi}}]{milais}%
  \BibitemOpen
  \bibfield  {author} {\bibinfo {author} {\bibfnamefont {G.}~\bibnamefont
  {Milano}}, \bibinfo {author} {\bibfnamefont {G.}~\bibnamefont {Pedretti}},
  \bibinfo {author} {\bibfnamefont {M.}~\bibnamefont {Fretto}}, \bibinfo
  {author} {\bibfnamefont {L.}~\bibnamefont {Boarino}}, \bibinfo {author}
  {\bibfnamefont {F.}~\bibnamefont {Benfenati}}, \bibinfo {author}
  {\bibfnamefont {D.}~\bibnamefont {Ielmini}}, \bibinfo {author} {\bibfnamefont
  {I.}~\bibnamefont {Valov}}, \ and\ \bibinfo {author} {\bibfnamefont
  {C.}~\bibnamefont {Ricciardi}},\ }\href {\doibase
  https://doi.org/10.1002/aisy.202000096} {\bibfield  {journal} {\bibinfo
  {journal} {Advanced Intelligent Systems}\ }\textbf {\bibinfo {volume} {2}},\
  \bibinfo {pages} {2000096} (\bibinfo {year} {2020}{\natexlab{b}})},\ \Eprint
  {http://arxiv.org/abs/https://onlinelibrary.wiley.com/doi/pdf/10.1002/aisy.202000096}
  {https://onlinelibrary.wiley.com/doi/pdf/10.1002/aisy.202000096} \BibitemShut
  {NoStop}%
\bibitem [{\citenamefont {Sheldon}\ and\ \citenamefont
  {Di~Ventra}(2017)}]{sheldonava}%
  \BibitemOpen
  \bibfield  {author} {\bibinfo {author} {\bibfnamefont {F.~C.}\ \bibnamefont
  {Sheldon}}\ and\ \bibinfo {author} {\bibfnamefont {M.}~\bibnamefont
  {Di~Ventra}},\ }\href {\doibase 10.1103/PhysRevE.95.012305} {\bibfield
  {journal} {\bibinfo  {journal} {Phys. Rev. E}\ }\textbf {\bibinfo {volume}
  {95}},\ \bibinfo {pages} {012305} (\bibinfo {year} {2017})}\BibitemShut
  {NoStop}%
\bibitem [{\citenamefont {Sattar}\ \emph {et~al.}(2013)\citenamefont {Sattar},
  \citenamefont {Fostner},\ and\ \citenamefont {Brown}}]{BrownPart}%
  \BibitemOpen
  \bibfield  {author} {\bibinfo {author} {\bibfnamefont {A.}~\bibnamefont
  {Sattar}}, \bibinfo {author} {\bibfnamefont {S.}~\bibnamefont {Fostner}}, \
  and\ \bibinfo {author} {\bibfnamefont {S.~A.}\ \bibnamefont {Brown}},\ }\href
  {\doibase 10.1103/PhysRevLett.111.136808} {\bibfield  {journal} {\bibinfo
  {journal} {Phys. Rev. Lett.}\ }\textbf {\bibinfo {volume} {111}},\ \bibinfo
  {pages} {136808} (\bibinfo {year} {2013})}\BibitemShut {NoStop}%
\bibitem [{\citenamefont {Daniels}\ \emph {et~al.}(2022)\citenamefont
  {Daniels}, \citenamefont {Mallinson}, \citenamefont {Heywood}, \citenamefont
  {Bones}, \citenamefont {Arnold},\ and\ \citenamefont
  {Brown}}]{DANIELS2022122}%
  \BibitemOpen
  \bibfield  {author} {\bibinfo {author} {\bibfnamefont {R.}~\bibnamefont
  {Daniels}}, \bibinfo {author} {\bibfnamefont {J.}~\bibnamefont {Mallinson}},
  \bibinfo {author} {\bibfnamefont {Z.}~\bibnamefont {Heywood}}, \bibinfo
  {author} {\bibfnamefont {P.}~\bibnamefont {Bones}}, \bibinfo {author}
  {\bibfnamefont {M.}~\bibnamefont {Arnold}}, \ and\ \bibinfo {author}
  {\bibfnamefont {S.}~\bibnamefont {Brown}},\ }\href {\doibase
  https://doi.org/10.1016/j.neunet.2022.07.001} {\bibfield  {journal} {\bibinfo
   {journal} {Neural Networks}\ }\textbf {\bibinfo {volume} {154}},\ \bibinfo
  {pages} {122} (\bibinfo {year} {2022})}\BibitemShut {NoStop}%
\bibitem [{\citenamefont {Pike}\ \emph {et~al.}(2020)\citenamefont {Pike},
  \citenamefont {Bose}, \citenamefont {Mallinson}, \citenamefont {Acharya},
  \citenamefont {Shirai}, \citenamefont {Galli}, \citenamefont {Weddell},
  \citenamefont {Bones}, \citenamefont {Arnold},\ and\ \citenamefont
  {Brown}}]{brownxx}%
  \BibitemOpen
  \bibfield  {author} {\bibinfo {author} {\bibfnamefont {M.~D.}\ \bibnamefont
  {Pike}}, \bibinfo {author} {\bibfnamefont {S.~K.}\ \bibnamefont {Bose}},
  \bibinfo {author} {\bibfnamefont {J.~B.}\ \bibnamefont {Mallinson}}, \bibinfo
  {author} {\bibfnamefont {S.~K.}\ \bibnamefont {Acharya}}, \bibinfo {author}
  {\bibfnamefont {S.}~\bibnamefont {Shirai}}, \bibinfo {author} {\bibfnamefont
  {E.}~\bibnamefont {Galli}}, \bibinfo {author} {\bibfnamefont {S.~J.}\
  \bibnamefont {Weddell}}, \bibinfo {author} {\bibfnamefont {P.~J.}\
  \bibnamefont {Bones}}, \bibinfo {author} {\bibfnamefont {M.~D.}\ \bibnamefont
  {Arnold}}, \ and\ \bibinfo {author} {\bibfnamefont {S.~A.}\ \bibnamefont
  {Brown}},\ }\bibfield  {booktitle} {\emph {\bibinfo {booktitle} {Nano
  Letters}},\ }\href {\doibase 10.1021/acs.nanolett.0c01096} {\bibfield
  {journal} {\bibinfo  {journal} {Nano Letters}\ }\textbf {\bibinfo {volume}
  {20}},\ \bibinfo {pages} {3935} (\bibinfo {year} {2020})}\BibitemShut
  {NoStop}%
\bibitem [{\citenamefont {Nakagaki}\ \emph {et~al.}(2000)\citenamefont
  {Nakagaki}, \citenamefont {Yamada},\ and\ \citenamefont {{A.
  T{\'{o}}th}}}]{Nakagaki2000}%
  \BibitemOpen
  \bibfield  {author} {\bibinfo {author} {\bibfnamefont {T.}~\bibnamefont
  {Nakagaki}}, \bibinfo {author} {\bibfnamefont {H.}~\bibnamefont {Yamada}}, \
  and\ \bibinfo {author} {\bibnamefont {{A. T{\'{o}}th}}},\ }\href {\doibase
  10.1038/35035159} {\bibfield  {journal} {\bibinfo  {journal} {Nature}\
  }\textbf {\bibinfo {volume} {407}},\ \bibinfo {pages} {470} (\bibinfo {year}
  {2000})}\BibitemShut {NoStop}%
\bibitem [{\citenamefont {Tero}\ \emph {et~al.}(2007)\citenamefont {Tero},
  \citenamefont {Kobayashi},\ and\ \citenamefont {Nakagaki}}]{Tero_2007}%
  \BibitemOpen
  \bibfield  {author} {\bibinfo {author} {\bibfnamefont {A.}~\bibnamefont
  {Tero}}, \bibinfo {author} {\bibfnamefont {T.}~\bibnamefont {Kobayashi}}, \
  and\ \bibinfo {author} {\bibfnamefont {T.}~\bibnamefont {Nakagaki}},\ }\href
  {\doibase 10.1016/j.jtbi.2006.07.015} {\bibfield  {journal} {\bibinfo
  {journal} {Journal of Theoretical Biology}\ }\textbf {\bibinfo {volume}
  {244}},\ \bibinfo {pages} {553} (\bibinfo {year} {2007})}\BibitemShut
  {NoStop}%
\bibitem [{\citenamefont {Alim}\ \emph {et~al.}(2017)\citenamefont {Alim},
  \citenamefont {Andrew}, \citenamefont {Pringle},\ and\ \citenamefont
  {Brenner}}]{pnasphys}%
  \BibitemOpen
  \bibfield  {author} {\bibinfo {author} {\bibfnamefont {K.}~\bibnamefont
  {Alim}}, \bibinfo {author} {\bibfnamefont {N.}~\bibnamefont {Andrew}},
  \bibinfo {author} {\bibfnamefont {A.}~\bibnamefont {Pringle}}, \ and\
  \bibinfo {author} {\bibfnamefont {M.~P.}\ \bibnamefont {Brenner}},\ }\href
  {\doibase 10.1073/pnas.1618114114} {\bibfield  {journal} {\bibinfo  {journal}
  {Proceedings of the National Academy of Sciences}\ }\textbf {\bibinfo
  {volume} {114}},\ \bibinfo {pages} {5136} (\bibinfo {year} {2017})},\ \Eprint
  {http://arxiv.org/abs/https://www.pnas.org/doi/pdf/10.1073/pnas.1618114114}
  {https://www.pnas.org/doi/pdf/10.1073/pnas.1618114114} \BibitemShut {NoStop}%
\bibitem [{\citenamefont {Adamatzky}(2012)}]{adamatzky}%
  \BibitemOpen
  \bibfield  {author} {\bibinfo {author} {\bibfnamefont {A.}~\bibnamefont
  {Adamatzky}},\ }\href {\doibase 10.1142/8482} {\emph {\bibinfo {title}
  {Bioevaluation of World Transport Networks}}}\ (\bibinfo  {publisher} {WORLD
  SCIENTIFIC},\ \bibinfo {year} {2012})\ \Eprint
  {http://arxiv.org/abs/https://www.worldscientific.com/doi/pdf/10.1142/8482}
  {https://www.worldscientific.com/doi/pdf/10.1142/8482} \BibitemShut {NoStop}%
\bibitem [{\citenamefont {Bonifaci}\ \emph {et~al.}(2012)\citenamefont
  {Bonifaci}, \citenamefont {Mehlhorn},\ and\ \citenamefont
  {Varma}}]{BONIFACI2012121}%
  \BibitemOpen
  \bibfield  {author} {\bibinfo {author} {\bibfnamefont {V.}~\bibnamefont
  {Bonifaci}}, \bibinfo {author} {\bibfnamefont {K.}~\bibnamefont {Mehlhorn}},
  \ and\ \bibinfo {author} {\bibfnamefont {G.}~\bibnamefont {Varma}},\ }\href
  {\doibase https://doi.org/10.1016/j.jtbi.2012.06.017} {\bibfield  {journal}
  {\bibinfo  {journal} {Journal of Theoretical Biology}\ }\textbf {\bibinfo
  {volume} {309}},\ \bibinfo {pages} {121} (\bibinfo {year}
  {2012})}\BibitemShut {NoStop}%
\bibitem [{\citenamefont {Milano}\ \emph
  {et~al.}(2022{\natexlab{c}})\citenamefont {Milano}, \citenamefont {Aono},
  \citenamefont {Boarino}, \citenamefont {Celano}, \citenamefont {Hasegawa},
  \citenamefont {Kozicki}, \citenamefont {Majumdar}, \citenamefont {Menghini},
  \citenamefont {Miranda}, \citenamefont {Ricciardi}, \citenamefont
  {Tappertzhofen}, \citenamefont {Terabe},\ and\ \citenamefont
  {Valov}}]{quantumcond}%
  \BibitemOpen
  \bibfield  {author} {\bibinfo {author} {\bibfnamefont {G.}~\bibnamefont
  {Milano}}, \bibinfo {author} {\bibfnamefont {M.}~\bibnamefont {Aono}},
  \bibinfo {author} {\bibfnamefont {L.}~\bibnamefont {Boarino}}, \bibinfo
  {author} {\bibfnamefont {U.}~\bibnamefont {Celano}}, \bibinfo {author}
  {\bibfnamefont {T.}~\bibnamefont {Hasegawa}}, \bibinfo {author}
  {\bibfnamefont {M.}~\bibnamefont {Kozicki}}, \bibinfo {author} {\bibfnamefont
  {S.}~\bibnamefont {Majumdar}}, \bibinfo {author} {\bibfnamefont
  {M.}~\bibnamefont {Menghini}}, \bibinfo {author} {\bibfnamefont
  {E.}~\bibnamefont {Miranda}}, \bibinfo {author} {\bibfnamefont
  {C.}~\bibnamefont {Ricciardi}}, \bibinfo {author} {\bibfnamefont
  {S.}~\bibnamefont {Tappertzhofen}}, \bibinfo {author} {\bibfnamefont
  {K.}~\bibnamefont {Terabe}}, \ and\ \bibinfo {author} {\bibfnamefont
  {I.}~\bibnamefont {Valov}},\ }\href {\doibase
  https://doi.org/10.1002/adma.202201248} {\bibfield  {journal} {\bibinfo
  {journal} {Advanced Materials}\ }\textbf {\bibinfo {volume} {34}},\ \bibinfo
  {pages} {2201248} (\bibinfo {year} {2022}{\natexlab{c}})},\ \Eprint
  {http://arxiv.org/abs/https://onlinelibrary.wiley.com/doi/pdf/10.1002/adma.202201248}
  {https://onlinelibrary.wiley.com/doi/pdf/10.1002/adma.202201248} \BibitemShut
  {NoStop}%
\bibitem [{\citenamefont {Wang}\ \emph {et~al.}(2019)\citenamefont {Wang},
  \citenamefont {Wang}, \citenamefont {Ambrosi}, \citenamefont {Bricalli},
  \citenamefont {Laudato}, \citenamefont {Sun}, \citenamefont {Chen},\ and\
  \citenamefont {Ielmini}}]{ielminisd}%
  \BibitemOpen
  \bibfield  {author} {\bibinfo {author} {\bibfnamefont {W.}~\bibnamefont
  {Wang}}, \bibinfo {author} {\bibfnamefont {M.}~\bibnamefont {Wang}}, \bibinfo
  {author} {\bibfnamefont {E.}~\bibnamefont {Ambrosi}}, \bibinfo {author}
  {\bibfnamefont {A.}~\bibnamefont {Bricalli}}, \bibinfo {author}
  {\bibfnamefont {M.}~\bibnamefont {Laudato}}, \bibinfo {author} {\bibfnamefont
  {Z.}~\bibnamefont {Sun}}, \bibinfo {author} {\bibfnamefont {X.}~\bibnamefont
  {Chen}}, \ and\ \bibinfo {author} {\bibfnamefont {D.}~\bibnamefont
  {Ielmini}},\ }\href@noop {} {\bibfield  {journal} {\bibinfo  {journal}
  {Nature Communications}\ }\textbf {\bibinfo {volume} {10}} (\bibinfo {year}
  {2019})}\BibitemShut {NoStop}%
\end{thebibliography}%
\ifpnas
\else
\appendix
\clearpage
\onecolumngrid
\begin{center}
\textbf{\Large Supplementary Information}
\end{center}


\section{Nodal approach}

An example of the two-probe conductance measurements used in a typical nanowire experiment is shown in Fig. \ref{fig:exp}.
\begin{figure*}[h!]
    \centering
    \includegraphics{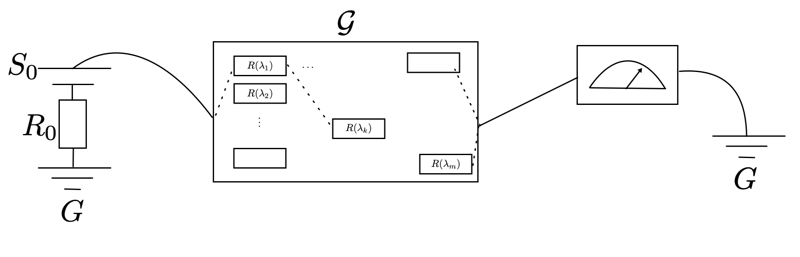}
    \caption{Experiment setup. The graph of conductances is represented by $\mathcal G$, while the external voltage generator (or current generator) is attached to nodes of the circuit. This is the typical setup for a two-probe conductance measurement.}
    \label{fig:exp}
\end{figure*}

The nodal analysis is based on a graph theoretical representation.
Given the memristive circuit graph, we introduce the directed incidence matrix $B$ of the graph. We assume a situation similar to the one of Fig. 1.

In the nodal approach, we begin with the current space, which is associated with the space of current configurations. If no external current is present, we can write
\begin{eqnarray}
B\vec i=\vec j_{ext}
\end{eqnarray}
where $\vec i$ has the cardinality of the edges, while $\vec j_{ext}$ has the cardinality of the nodes, and are the currents injected into that particular node. We will get back to this.
For every node, we assume that there are no internal voltage generators, and thus
\begin{eqnarray}
   {\bm G} \Delta \vec v=\vec i, 
\end{eqnarray}
where ${\bm G}$ is the conductance matrix. The potential drop on each edge can be written, using the potentials at the nodes, as 
\begin{eqnarray}
\Delta \vec v=B^t \vec v.
\end{eqnarray}
Note that $\Delta \vec v$ and $\vec v$ are vectors of different cardinalities, e.g. $\Delta \vec v$ has the cardinality of the number of edges, while $\vec v$ has the cardinality of the number of nodes. We can rewrite
\begin{eqnarray}
B{\bm G} \Delta \vec v=B {\bm G}B^t \vec v=B\vec i=\vec j_{ext}
\end{eqnarray}
Now we can write, then
\begin{eqnarray}
 \vec v=(B{\bm G}B^t)^{-1} \vec j_{ext}
\end{eqnarray}

\subsection{Proof of Lemma 1}
To get back to the voltage drops, we now apply $B^t$ on the left, and get
\begin{eqnarray}
 \Delta \vec v=B^t \vec v=B^t(B{\bm G}B^t)^{-1} \vec j_{ext}
 \label{eq:vapp1}
\end{eqnarray}
Now note that for this equation to make sense, we must have a current $j_{N+1}$ entering into node, say, $n_1$ and $-j_{N+1}$ on node $n_2\neq n_1$. This equation makes sense if we our external device is in current lock mode. If instead we apply a certain voltage $v_{N+1}/2$ at node $n_1$, and $-v_{N+1}$/2 at node $n_2$, we must have, given an external conductance $G_{N+1}=R_0^{-1}$, that
\begin{eqnarray}
 G_{N+1} v_{N+1}= j_{N+1}.
\end{eqnarray} 

We can then introduce an extra edge into the graph, which means increasing the number of columns of $B$, with a fictitious edge between $n_1$ and $n_2$ (say, the first row) with conductance $G_{N+1}$, and enlarging also the matrix ${\bm G}$ to contain $G_{N+1}$ in position $\tilde G_{(N+1)(N+1)}$. We call these extended matrices $\tilde {\bm G}$ and $\tilde B$.  Then, $(\Delta \vec v)_0$ must be $v_{N+1}$, and we must have
\begin{eqnarray}
\vec j_{ext}=G_{N+1} \tilde B \vec v_s.
\end{eqnarray}
As a matter of fact, the equation above is the one for a circuit in which we have a voltage generator $v_{N+1}$ in series with a conductance $G_{N+1}$. Since on has to fix $G_{N+1} v_{N+1}=j_{N+1}$, we need to evaluate the pseudo-inverse of the matrix $\tilde B {\bm \tilde G} \tilde B$ and then send $G_{N+1}\rightarrow 0$ to reobtain the same result as $\Delta \vec v=B^t(\tilde B {\bm \tilde G} \tilde B)^{-1} \vec j_{ext}$. We also checked numerically that this was true on some simple and some non-trivial circuits. Alternatively, we can assume that our external source is a voltage generator, and keep $G_{N+1}$ fixed, and if interpreted as a memristor, thus $g_{N+1}$ fixed.

In the equation above, $v_s$ contains $v_{N+1}$ in position $0$. With this formalism, we then have
\begin{eqnarray}
 \Delta \vec v=G_{N+1} {\tilde B}^t(\tilde B\tilde {\bm G}\tilde {B}^t)^{-1} \tilde B \vec v_s.
 \label{eq:vapp2}
\end{eqnarray}
We see then that we have proven
 \textbf{Lemma 1} of the main text.
 
\subsection{Effective Conductance definition}
For consistency, it is important to note that in the global circuit, $\Delta \vec v$ above satisfies the Kirchoff Voltage Law. This can be written, in terms of the loop matrix $A$ of the generalized circuit, as $\tilde A\Delta \vec v=0$. However, above this can be seen to be true because of Tellegen's theorem, which ensures that $\tilde A{\tilde B}^t=0$.

At this point, this is the equation for the voltage drops on the memristors in the nodal analysis.
Note that we now study
\begin{eqnarray}
 \tilde {\bm G}/G_0=\begin{pmatrix}
 G(g_1)/G_{N+1} & 0 & \cdots &0 \\
 0 & \ddots & 0  &0\\
 0 & 0 & G(g_N)/G_{N+1} & 0 \\
 0 & 0 & 0 & 1
 \end{pmatrix}
\end{eqnarray}
where we assume that $G(g_1)$ satisfies the equations above. It follows that we can write
\begin{eqnarray}
 {\bm {\tilde G}}/G_{min}=I+\Delta G
\end{eqnarray}
where
\begin{eqnarray}
 \Delta G=\begin{pmatrix}
  G_1(g_1)/G_{min} -1 & 0 & \cdots &0  \\
 0 & 0& 0 & 0\\
 0 & 0 & \vdots & 0\\
 0 & \cdots & 0& G_{N+1}/G_{min}
 \end{pmatrix}
\end{eqnarray}
Here, we start to make the assumption that the conductances ${\bm G}_{ii}=G_i(g_i)$ depend on a parameter $g_i$.
 and we can write
\begin{eqnarray}
 \Delta \vec v= {\tilde B}^t(\tilde B\tilde {B}^t+\tilde B\Delta  G\tilde {B}^t)^{-1} \tilde B \vec v_s.
 \label{eq:split}
\end{eqnarray}
We now want to write the equation above in terms of the projector matrix $\Omega_{\tilde B}={\tilde B}^t(\tilde B\tilde {B}^t)^{-1} \tilde B$.
To do this, we assume that $\tilde B\tilde {B}^t$ is invertible, which can be achieved by removing one node (this is a little technical, but alternatively it is also sufficient to replace $\ ^{-1}$ with the numerical pseudoinverse.

Let us now briefly comment on how to obtain the effective voltage of the whole device. 
In our setup, we have a 1-port device, in which we inject a current $j_{N+1}$, in series with another conductance $G_{N+1}$. Then, we must have
\begin{eqnarray}
\Delta v_{N+1}=\Delta v_{device}=\frac{j_{ext}}{G_{eff}}
\end{eqnarray}
because of KVL.
It follows that we must have
\begin{eqnarray}
\Delta v_{N+1}= G_{eff}^{-1} j_{N+1},
\end{eqnarray}
where we write $G_{eff}$ as the effective conductance. Then, we must have that
\begin{eqnarray}
G_{eff}=\frac{j_{N+1}}{\Delta v_{N+1}}.\label{eq:effcond}
\end{eqnarray}
We will use later the formula above to obtain the effective conductance in this setup.

\subsection{Network equation for the conductances}
The device $i$ parameter $g_i$ satisfies the evolution equation
\begin{eqnarray}
    \frac{dg_i}{dt}&=&\eta_{P}^i(\Delta v_i)(1-g_i)-\eta_{D}^i(\Delta v_i) g_i \nonumber \\
    &=&\eta_{P}^i(\Delta v_i)-\Big(\eta_{P}^i(\Delta v_i)+\eta_{D}^i(\Delta v_i)\Big)g_i
\end{eqnarray}
where
\begin{eqnarray}
    \eta_{P,D}^i(V)=\kappa_{P0,D0}^i e^{\eta_{P,D}^i V}
\end{eqnarray}
and in particular we also use, assuming that all the elements are homogeneous
\begin{eqnarray}
    G(g_i)/G_{min}&=&G_{min}/G_{min}\left(1+\frac{(G_{MAX}-G_{min})}{G_{min}} g_i\right) \nonumber \\
    &\equiv& 1+\chi g_i
\end{eqnarray}
Note then that we can write
\begin{eqnarray}
 \Delta \vec v= {\tilde B}^t(\tilde B\tilde {B}^t+\chi \tilde B \bm{g}\tilde {B}^t)^{-1} \tilde B \vec v_s.
\end{eqnarray}
where we introduced the matrix  ${\bm g}_{ii}=g_i$.

We now assume that the parameters are homogeneous, and after a brief calculation, we get
\begin{eqnarray}
 \Delta \vec v= \frac{G_{N+1}}{G_{min}}{\tilde B}^t(\tilde B\tilde {B}^t+\chi\tilde B \boldsymbol{  g}\tilde {B}^t)^{-1} \tilde B \vec v_s.
\end{eqnarray}
Then, we use the identity
\begin{eqnarray}
 {\tilde B}^t(\tilde B\tilde {B}^t+\chi \tilde B  {\bm g}\tilde {B}^t)^{-1} \tilde B=(I+\chi \Omega_{\tilde B} {\bm g})^{-1} \Omega_{\tilde B} \
\end{eqnarray}
from which we get
\begin{eqnarray}
 \Delta \vec v= \frac{G_{N+1}}{G_{min}} (I+\chi \Omega_{\tilde B} {\bm g})^{-1} \Omega_{\tilde B}\vec v_s.
 \label{eq:vapp}
\end{eqnarray}
We now we make the assumptions about the evolution of the parameters $g_i$ for $i>0$. 
 
We can now put together the equations, obtaining
\begin{eqnarray}
    \frac{dg_i}{dt}&=&\eta_{P}^i(\Delta v_i)(1-g_i)-\eta_{D}^i(\Delta v_i) g_i \nonumber \\
    &=&\eta_{P}^i(\Delta v_i)-\Big(\eta_{P}^i(\Delta v_i)+\eta_{D}^i(\Delta v_i)\Big)g_i
\end{eqnarray}
where $\Delta \vec v_i=\frac{G_{N+1}}{G_{min}} \sum_{j=1}^N\Big((I+\chi \Omega_{\tilde B} {\bm g})^{-1} \Omega_{\tilde B}\Big)_{ij}(\vec v_s)_j$. Of course, the formula above is true if we had that also the conductance on the source was a memristor, which of course is false as it is a fixed conductance. Since we used essentially that
\begin{eqnarray}
 \Delta G=\begin{pmatrix}
 G_1(g_1)/G_{min} -1 & 0 & \cdots & 0 \\
 0 & \ddots & 0  & 0\\
 0 & 0 & \ddots  & 0\\
 0 & 0& \cdots & G_{N+1}(g_{N+1})/G_{min} -1\\
 \end{pmatrix}
\end{eqnarray}
we note that we need to be careful here, as $g_{N+1}$ is special. We need to set artificially $g_{N+1}$ to satisfy $1+\chi g_{N+1}=G_{N+1}$, or
\begin{eqnarray}
    g_{N+1}=\frac{(G_{N+1}-1)G_{min}}{G_{MAX}-G_{min}}
\end{eqnarray}

The analysis above is for a voltage generator $v_{N+1}$ in series with a conductance $G_{N+1}$. If we want to fix the external current, then we need to fix $G_{N+1} \tilde v_s$, and take the limit $G_{N+1}\rightarrow 0$. Then, in this case we must have $\lim_{G_{N+1}\rightarrow 0}g_{N+1}=-\frac{1}{\chi}$. In any case, we can use this formalism pretending that $g_{N+1}$ is a memristor, and fixing its value a posteriori depending on the situation.

\subsection{Matrix inverse in detail - Proof of Corollary 1}
From now on, we will simply write $\Omega_{\tilde B}$ as $\Omega$, to avoid overburdening the notation.
We want to make more explicit the matrix inverse, to see if we can improve the matrix inverse mean field.
As we saw, we  have
\begin{eqnarray}
\Delta \vec v=\frac{G_{N+1}}{G_{min}} (I+\chi \Omega\bm{g})^{-1}\Omega \vec v_s
\end{eqnarray}
and we want to write this expression explicitly in terms of the $g_i$'s from the junctions. For this purpose, 
let us write $\Omega$ in block diagonal form
\begin{eqnarray}
\Omega=\begin{pmatrix}
\tilde \Omega & \vec \Omega\\
\vec \Omega^t & \Omega_{N+1}
\end{pmatrix}
\end{eqnarray}
and $\bm{g}=\text{diag}(\vec g,g_{N+1})$. 
Let us call $\tilde{\bm{g}}=\text{diag}(\vec g)$.
We can write
\begin{eqnarray}
(I+\chi \Omega \bm{g})^{-1}&=&\begin{pmatrix}
I+\chi \tilde \Omega \tilde {\bm{g}} & \chi g_{N+1} \vec \Omega \\
\chi (\tilde {\bm{g}} \vec \Omega)^t & 1+\chi g_{N+1} \Omega_{N+1}
\end{pmatrix}^{-1}
\end{eqnarray}
We now use the matrix block inverse identity
\begin{eqnarray}
\begin{pmatrix}
A & B \\
C & D
\end{pmatrix}^{-1}=\begin{pmatrix}
Q_{11} & \vec Q_{12}\\
\vec Q_{21}^t & Q_{22}
\end{pmatrix}=\begin{pmatrix}
A^{-1}+A^{-1} B(D-C A^{-1} B)^{-1} C A^{-1} & -A^{-1} B (D-CA^{-1 }B)^{-1} \\
-(D-C A^{-1} B)^{-1} C A^{-1} & (D-CA^{-1} B)^{-1}
\end{pmatrix}
\end{eqnarray}
We have $A=I+\chi \tilde \Omega \tilde {\bm{g}}$, which we stress is a $N\times N$ matrix.
Let us now focus on $q_0=D-CA^{-1} B$.
This quantity is a scalar, given by
\begin{eqnarray}
q_0&=&1+\chi g_{N+1}( \Omega_{N+1}-\chi \vec \Omega^{t}\tilde{\bm{g}} A^{-1} \vec \Omega).
\label{eq:q0}
\end{eqnarray}
We define the rank-1 matrix $$f_0=\chi^2 g_{N+1} (\vec \Omega) \otimes \big({\tilde {\bm{ g}}}\vec \Omega\big)^t.$$
First, we have $$Q_{22}=q_0^{-1}.$$
We have
\begin{eqnarray}
Q_{11}=(I+\chi \tilde \Omega \tilde{\bm{g}})^{-1} +q_0^{-1} (I+\chi \tilde \Omega \tilde{\bm{g}})^{-1}f_0 (I+\chi \tilde \Omega \tilde{\bm{g}})^{-1}
\end{eqnarray}
and we get
\begin{eqnarray}
\vec {Q}_{12}=-\chi\frac{g_{N+1}}{q_0} (I+\chi \tilde \Omega \tilde{\bm{g}})^{-1} \vec \Omega,
\end{eqnarray}
while
\begin{eqnarray}
\vec Q_{21}^t=-\chi \frac{1}{q_0} \vec \Omega^t\tilde{\bm{g}}(I+\chi \tilde \Omega \tilde{\bm{g}})^{-1}  .
\end{eqnarray}
Let us now focus on $\Omega \vec v_s$. We write
\begin{eqnarray}
\vec v_s=\begin{pmatrix}
\tilde v\\
v_{N+1}
\end{pmatrix}
\end{eqnarray}
thus we have
\begin{eqnarray}
\Omega \vec v_s=\begin{pmatrix}
\tilde \Omega & \vec \Omega\\
\vec \Omega^t & \Omega_{N+1}
\end{pmatrix}\begin{pmatrix}
\tilde v\\
v_{N+1}
\end{pmatrix}=\begin{pmatrix}
\tilde \Omega \tilde v+v_{N+1} \vec \Omega\\
\Omega_{N+1} v_{N+1}+\vec \Omega^t \tilde v
\end{pmatrix}=\begin{pmatrix}
\vec v_a\\
v_b
\end{pmatrix}
\end{eqnarray}
and thus we get
\begin{eqnarray}
(I+\chi \Omega \bm{g})^{-1} \Omega \vec v_s=\begin{pmatrix}
Q_{11} & \vec Q_{12}\\
\vec Q_{21}^t & Q_{22}
\end{pmatrix}\begin{pmatrix}
\vec v_a\\
v_b
\end{pmatrix}=\begin{pmatrix}
R_{1}\\
R_{2} 
\end{pmatrix}
\end{eqnarray}
where
\begin{eqnarray}
R_{1}&=&Q_{11} \vec v_a+ \vec Q_{12} v_b= Q_{11}(\tilde \Omega \tilde v+ v_{N+1} \vec \Omega)+(\Omega_{N+1} v_{N+1}+ \vec \Omega^t \tilde v_a)\vec Q_{12}\\
R_{2} &=& \vec Q_{21}^t \vec v_a+Q_{22} v_b=\vec Q_{21}^t (\tilde \Omega \tilde v+ v_{N+1} \vec \Omega)+(\Omega_{N+1}v_{N+1}+\vec \Omega^t \tilde v)Q_{22}
\end{eqnarray}
If we now assume that $\tilde v=0$, we get
\begin{eqnarray}
R_{1}&=& v_{N+1}(Q_{11} \vec \Omega+\Omega_{N+1} \vec Q_{12})\\
R_{2} &=&   v_{N+1} (\vec Q_{21}^t\vec \Omega+\Omega_{N+1}Q_{22})
\end{eqnarray}
and thus we get
\begin{eqnarray}
\Delta\vec v_{all}&= &\frac{G_{N+1}}{G_{min}}\begin{pmatrix}
R_{1}\\
R_{2} 
\end{pmatrix}\nonumber \\
&=&v_{N+1}\frac{G_{N+1}}{G_{min}}\begin{pmatrix}
Q_{11} \vec \Omega+\Omega_{N+1} \vec Q_{12}\\
\vec Q_{21}^t\vec \Omega+\Omega_{N+1}Q_{22}
\end{pmatrix}\\
&=&\frac{G_{N+1}v_{N+1}}{G_{min}}\begin{pmatrix}
\Big((I+\chi \tilde \Omega \tilde{\bm{g}})^{-1} +q_0^{-1} (I+\chi \tilde \Omega \tilde{\bm{g}})^{-1}f_0 (I+\chi \tilde \Omega \tilde{\bm{g}})^{-1}\Big)\vec \Omega-g_{N+1}\chi q_0^{-1}\Omega_{N+1}(I+\chi \tilde \Omega \tilde{\bm{g}})^{-1} \vec \Omega\nonumber  \\
-\frac{1}{q_0}(\chi\vec \Omega^t \tilde{\bm{g}}(I+\chi \tilde\Omega\tilde{\bm{g}})^{-1}\vec \Omega-\Omega_{N+1})
\end{pmatrix}\\
&=& \frac{G_{N+1}v_{N+1}}{G_{min}}\begin{pmatrix}
\Big(I +\frac{ (I+\chi \tilde \Omega \tilde{\bm{g}})^{-1}f_0-\chi\Omega_{N+1} g_{N+1} I }{1+\chi g_{N+1}( \Omega_{N+1}- \chi\vec \Omega^{t}\tilde{\bm{g}} A^{-1} \vec \Omega)} \Big)(I+\chi \tilde \Omega \tilde{\bm{g}})^{-1}\vec \Omega\\
\frac{\eta}{q_0}
\end{pmatrix}
\end{eqnarray}
where we have 
called
\begin{eqnarray}
\rho= \Omega_{N+1}- \chi\vec \Omega^{t}\tilde{\bm{g}} (I+\chi \tilde \Omega \tilde{\bm{g}})^{-1} \vec \Omega
\label{eq:etasm}
\end{eqnarray}

Note that if $G_{N+1}\rightarrow 0$, we have $g_{N+1}\rightarrow -\frac{1}{\chi}$.
The formula above does not have any approximations.

The voltage drop on the devices is the vector of internal voltage drops, given by 
\begin{eqnarray}
\Delta \vec v_{int}&=&\frac{G_{N+1}v_{N+1}}{G_{min}}
\Big(I +\frac{ (I+\chi \tilde \Omega \tilde{\bm{g}})^{-1}f_0-\chi\Omega_{N+1} g_{N+1} I }{1+\chi g_{N+1}( \Omega_{N+1}- \chi\vec \Omega^{t}\tilde{\bm{g}} (I+\chi \tilde \Omega \tilde{\bm{g}})^{-1} \vec \Omega)} \Big)(I+\chi \tilde \Omega \tilde{\bm{g}})^{-1}\vec \Omega\nonumber \\
&=&\frac{G_{N+1}v_{N+1}}{G_{min}}
\Big(I +\frac{ g_{N+1}\chi^2(I+\chi \tilde \Omega \tilde{\bm{g}})^{-1}  (\vec \Omega) \otimes \big({\tilde {\bm{ g}}}\vec \Omega\big)^t-\chi\Omega_{N+1} g_{N+1} I }{1+\chi g_{N+1}( \Omega_{N+1}- \chi\vec \Omega^{t}\tilde{\bm{g}} (I+\chi \tilde \Omega \tilde{\bm{g}})^{-1} \vec \Omega)} \Big)(I+\chi \tilde \Omega \tilde{\bm{g}})^{-1}\vec \Omega\nonumber \\
&=&\frac{G_{N+1}v_{N+1}}{G_{min}}
\Big(I +g_{N+1}\chi\frac{ \chi \vec \Omega^t{\tilde {\bm{ g}}}(I+\chi \tilde \Omega \tilde{\bm{g}})^{-1}  \vec \Omega -\Omega_{N+1}   }{1+\chi g_{N+1}( \Omega_{N+1}- \chi\vec \Omega^{t}\tilde{\bm{g}} (I+\chi \tilde \Omega \tilde{\bm{g}})^{-1} \vec \Omega)} \Big)(I+\chi \tilde \Omega \tilde{\bm{g}})^{-1}\vec \Omega\nonumber \\
\end{eqnarray}
and using the fact that $g_{N+1}\approx -1/\chi$, we have

\begin{eqnarray}
\Delta \vec v_{int}
&=&\frac{G_{N+1}v_{N+1}}{G_{min}}
\Big(1 +\frac{ \rho}{1-\rho} \Big)(I+\chi \tilde \Omega \tilde{\bm{g}})^{-1}\vec \Omega\nonumber \\
&=&\frac{1}{1-\rho}\frac{G_{N+1}v_{N+1}}{G_{min}}
(I+\chi \tilde \Omega \tilde{\bm{g}})^{-1}\vec \Omega
\label{eq:vexact}
\end{eqnarray}

We see then that the statement above is the proof of \textbf{Corollary 1}.

\subsection{Properties of projector operator}
Note that $\tilde \Omega$ is not a projector operator in general, unlike $\Omega$. We have however
\begin{eqnarray}
\begin{pmatrix}
\tilde \Omega & \vec \Omega\\
\vec \Omega^t & \Omega_{N+1}
\end{pmatrix}\begin{pmatrix}
\tilde \Omega & \vec \Omega\\
\vec \Omega^t & \Omega_{N+1}
\end{pmatrix}=\Omega^2=\Omega=\begin{pmatrix}
\tilde \Omega & \vec \Omega\\
\vec \Omega^t & \Omega_{N+1}
\end{pmatrix}
\end{eqnarray}
and thus
\begin{eqnarray}
\begin{pmatrix}
\tilde \Omega & \vec \Omega\\
\vec \Omega^t & \Omega_{N+1}
\end{pmatrix}\begin{pmatrix}
\tilde \Omega & \vec \Omega\\
\vec \Omega^t & \Omega_{N+1}
\end{pmatrix}=\begin{pmatrix}
\Omega^2+\vec \Omega\otimes \vec \Omega^t & (\tilde \Omega+\Omega_{N+1}) \vec \Omega \\
\vec \Omega^t( \tilde \Omega +\Omega_{N+1} I) & \vec \Omega^t\vec \Omega+\Omega_{N+1}^2
\end{pmatrix}=\begin{pmatrix}
\tilde \Omega & \vec \Omega\\
\vec \Omega^t & \Omega_{N+1}
\end{pmatrix}
\end{eqnarray}
From which we get
\begin{eqnarray}
\tilde \Omega^2+\vec \Omega\otimes \vec \Omega^t&=&\tilde \Omega\rightarrow \tilde \Omega^2=\tilde \Omega-\vec \Omega\otimes \vec \Omega^t\\
\tilde \Omega\vec\Omega&=&(1-\Omega_{N+1})\vec \Omega\\
\|\vec \Omega\|^2&=&\Omega_{N+1}(1-\Omega_{N+1})
\end{eqnarray}
we will use the formulae next.
\subsection{Proof of Corollary 2}
Let us now use the equations above to obtain results about the effective conductance of the whole device, where we can use eqn. (\ref{eq:effcond}), which we recall is
\begin{eqnarray}
G_{eff}=\frac{j_{N+1}}{\Delta v_{N+1}}.
\end{eqnarray}
We can then note that, from the equations above, we have
\begin{eqnarray}
\Delta v_{N+1}=\frac{G_{N+1} v_{N+1}}{G_{min}} \frac{\rho}{q_0}.
\end{eqnarray}
Similarly to the case of linear memristors, we then have explicit expressions for $q_0$ and $\rho$, eqn. (\ref{eq:q0}) and eqn. (\ref{eq:etasm2}) respectively.
Replacing, we have
\begin{eqnarray}
G_{eff}&=&\frac{j_{N+1}}{\frac{G_{N+1}v_{N+1}}{G_{min}}\frac{\rho}{q_0} }= G_{min}\frac{j_{N+1}}{G_{N+1}v_{N+1} }\frac{q_0}{\rho}\\
&=&G_{min}\frac{j_{N+1}}{G_{N+1}v_{N+1} }\frac{1+\chi g_{N+1}( \Omega_{N+1}-\chi \vec \Omega^{t}\tilde{\bm{g}} (I+\chi \tilde \Omega \tilde{\bm{g}})^{-1} \vec \Omega)}{\Omega_{N+1}- \chi\vec \Omega^{t}\tilde{\bm{g}} (I+\chi \tilde \Omega \tilde{\bm{g}})^{-1} \vec \Omega}\\
&=&G_{min}\frac{j_{N+1}}{G_{N+1}v_{N+1} }\frac{1+\chi g_{N+1} \rho}{\rho}.
\end{eqnarray}

We note that $G_{n+1}{v_{N+1}}=j_{N+1}$, and thus the expression simplifies to
\begin{eqnarray}
G_{eff}&=&G_{min}\frac{1+\chi g_{N+1}\rho}{\rho}=G_{min}\frac{1+\chi g_{N+1}( \Omega_{N+1}-\chi \vec \Omega^{t}\tilde{\bm{g}} (I+\chi \tilde \Omega \tilde{\bm{g}})^{-1} \vec \Omega)}{\Omega_{N+1}- \chi\vec \Omega^{t}\tilde{\bm{g}} (I+\chi \tilde \Omega \tilde{\bm{g}})^{-1} \vec \Omega}
\end{eqnarray}
which is an exact expression of the effective conductance of the whole device. Now, in the limit $G_{N+1}\rightarrow 0$, we have $g_{N+1}\rightarrow-\frac{1}{\chi}$. It follows that we have
\begin{eqnarray}
\lim_{G_{N+1}\rightarrow 0}G_{eff}=G_{min}\frac{1-\rho}{\rho}
\end{eqnarray}
We now see that the statement above is our proof of \textbf{Corollary 2} about the effective conductance.

\section{Matrix inverse approach and mean field theory voltage drops}
As mentioned, the issue is the matrix inverse given by
\begin{eqnarray}
(I+\chi \tilde \Omega \bm{g})^{-1}.
\end{eqnarray}
We then wonder if we could perform the approximation
\begin{eqnarray}
(I+\chi \tilde \Omega \bm{g})^{-1}\approx (I+\chi \tilde \Omega \langle g\rangle)^{-1}
\end{eqnarray}
where $\langle g\rangle$ is the minimum of a matrix norm of the form
\begin{eqnarray}
\langle g\rangle&=&\text{argmin}_{\langle g\rangle} \text{Tr}\big((I+\chi\tilde \Omega \bm{g})- (I+\chi \tilde \Omega\langle g\rangle) \big)^2 \\
&=&\chi^2\text{argmin}_{\langle g\rangle} \text{Tr}\big(\tilde \Omega \bm{g}-  \tilde \Omega\langle g\rangle \big)^2\\
&=& \chi^2\text{argmin}_{\langle g\rangle} \text{Tr}\big((\tilde \Omega \bm{g})^2+  (\tilde \Omega\langle g\rangle)^2 -2\langle g\rangle\tilde \Omega^2\bm{g}\big)
\end{eqnarray}
We can derive with respect to $\langle g\rangle$ the Frobenius norm, obtaining the maximization
\begin{eqnarray}
0&=&\partial_{\langle g\rangle}\chi^2\text{argmin}_{\langle g\rangle} \text{Tr}\big((\tilde \Omega \bm{g})^2+  (\tilde \Omega\langle g\rangle)^2 -2\langle g\rangle\tilde \Omega^2\bm{g}\big)\\
&=&2\chi^2\text{argmin}_{\langle g\rangle} \text{Tr}\big(\tilde \Omega^2\langle g\rangle -\tilde \Omega^2\bm{g}\big)\\
\end{eqnarray}
from which we get the value
\begin{eqnarray}
\langle g\rangle=\frac{\text{Tr}(\tilde \Omega^2\bm{g})}{\text{Tr}(\tilde \Omega^2)}
\end{eqnarray}
Note that $\tilde \Omega$ is not a projector, and thus $\tilde \Omega^2\neq \tilde \Omega$.
It follows that we have the norm-2 approximation to the inverse, given by
\begin{eqnarray}
(I+\chi \tilde \Omega)^{-1}\approx (I+\chi \tilde \Omega \frac{\text{Tr}(\tilde \Omega^2\bm{g})}{\text{Tr}(\tilde \Omega^2)})^{-1}\equiv (I+\chi \tilde \Omega \langle g\rangle)^{-1},
\end{eqnarray}
which we use next. Note that the approximation is consistent, e.g. that if $\bm{g}=\langle g\rangle I$, then $\langle g\rangle = \langle g\rangle$. Thus, the more homogeneous the memristors are the more precise the mean field theory becomes.

We can use the expression above for a mean field theory. If we use $\tilde{\bm{g}}=\langle g\rangle I$, where $\langle g\rangle$ has to be determined, and then we get
\begin{eqnarray}
\Delta \vec v\approx \frac{G_{N+1} v_{N+1}}{G_{min}(1-\rho)} (I+\chi \langle g\rangle \tilde \Omega)^{-1}\vec \Omega=\frac{G_{N+1} v_{N+1}}{G_{min}(1-\rho)} \frac{1}{1+\chi \langle g\rangle(1-\Omega_{N+1})}\vec \Omega
\end{eqnarray}
where $\langle g\rangle=\frac{\text{Tr}(\tilde\Omega \tilde{\bm{g}})}{\text{Tr}(\tilde \Omega)} $,
and for $\rho$ we have
\begin{eqnarray}
\rho&\approx& \Omega_{N+1}-\chi \frac{\langle g\rangle}{1+\chi\langle g\rangle(1-\Omega_{N+1})}\vec \Omega^t \vec \Omega\\
&=&\Omega_{N+1}-\chi \frac{\langle g\rangle}{1+\chi\langle g\rangle(1-\Omega_{N+1})}\Omega_{N+1}(1-\Omega_{N+1})\\
&=&\Omega_{N+1}\Big(1-\frac{\langle g\rangle\chi (1-\Omega_{N+1})}{1+\langle g\rangle\chi (1-\Omega_{N+1})}\Big)=\frac{\Omega_{N+1}}{{1+\langle g\rangle\chi (1-\Omega_{N+1})}}
\end{eqnarray}
and thus
\begin{eqnarray}
1-\rho&=&\frac{{1+\langle g\rangle\chi (1-\Omega_{N+1})}}{{1+\langle g\rangle\chi (1-\Omega_{N+1})}}-\frac{\Omega_{N+1}}{{1+\langle g\rangle\chi (1-\Omega_{N+1})}}\\
&=&\frac{1+\langle g\rangle \chi-\Omega_{N+1}(1+\langle g\rangle \chi)}{{1+\langle g\rangle\chi (1-\Omega_{N+1})}}=(1-\Omega_{N+1})\frac{1+\langle g\rangle\chi}{1+\langle g\rangle\chi(1-\Omega_{N+1})}
\end{eqnarray}
from which we then get
\begin{eqnarray}
\Delta \vec v\approx \Delta \vec v_{mft}=\frac{G_{N+1} v_{N+1}}{G_{min}(1-\Omega_{N+1})} \frac{1}{1+\langle g\rangle \chi}\vec \Omega
\label{eq:vmft}
\end{eqnarray}


At this point, we can also write a formula for the effective conductance in terms of the effective conductance, given by

\begin{eqnarray}
G_{eff}&=&G_{min}\frac{1-\rho}{\rho}=G_{min}\frac{(1-\Omega_{N+1})\frac{1+\langle g\rangle\chi}{1+\langle g\rangle\chi(1-\Omega_{N+1})}}{\frac{\Omega_{N+1}}{{1+\langle g\rangle\chi (1-\Omega_{N+1})}}}=\frac{1-\Omega_{N+1}}{\Omega_{N+1}}(1+\chi \langle g\rangle)\\
&=&\frac{1-\Omega_{N+1}}{\Omega_{N+1}}G(\langle g\rangle)
\end{eqnarray}
which is the equation for a global memristor.

\subsection{Mean field Miranda model fixed points}

Let us now insert this expression inside the Miranda model. We have 
\begin{eqnarray}
    \frac{dg_i}{dt}    &=&\kappa_{P0} e^{\eta_{P} |\Delta v_i|}-\Big(\kappa_{P0} e^{\eta_{P} |\Delta v_i|}+\kappa_{D0} e^{\eta_{D} |\Delta v_i|}\Big)g_i\nonumber \\
    &=&\kappa_{P0} e^{\eta_{P} \frac{G_{N+1} v_{N+1}}{G_{min}(1-\Omega_{N+1})} \frac{1}{1+\langle g\rangle \chi} |{\vec \Omega}_i|}-\Big(\kappa_{P0} e^{\eta_{P} \frac{G_{N+1} v_{N+1}}{G_{min}(1-\Omega_{N+1})} \frac{1}{1+\langle g\rangle \chi} |{\vec \Omega}_i|}+\kappa_{D0} e^{\eta_{D} \frac{G_{N+1} v_{N+1}}{G_{min}(1-\Omega_{N+1})} \frac{1}{1+\langle g\rangle \chi} |{\vec \Omega}_i|}\Big)g_i\nonumber 
\end{eqnarray}

In order to derive the mean field theory, where ${\vec \Omega}_i\sim c/N$ where $c$ is a constant. Then, we can write, multiplying by $a_i$ and summing
\begin{eqnarray}
\frac{d}{dt}\langle g\rangle=\kappa_{P0} e^{\eta_{P} \frac{G_{N+1} v_{N+1}}{G(\langle g\rangle)(1-\Omega_{N+1})}  \frac{c}{N}}-\Big(\kappa_{P0} e^{\eta_{P} \frac{G_{N+1} v_{N+1}}{G(\langle g\rangle)(1-\Omega_{N+1})}  \frac{c}{N}}+\kappa_{D0} e^{\eta_{D} \frac{G_{N+1} v_{N+1}}{G(\langle g\rangle)(1-\Omega_{N+1})}  \frac{c}{N}}\Big)\langle g\rangle
\end{eqnarray}

We can rewrite the expression above as
\begin{eqnarray}
\frac{d}{dt}\langle g\rangle=\kappa_{P0} e^{\frac{f\eta_{P} }{1+\chi \langle g\rangle}}  -\Big(\kappa_{P0} e^{\frac{f\eta_{P} }{1+\chi \langle g\rangle}}+\kappa_{D0} e^{ \frac{f\eta_{D} }{1+\chi \langle g\rangle}}\Big)\langle g\rangle=-\partial_{\langle g\rangle}V(\langle g\rangle).
\label{eq:mftd}
\end{eqnarray}
where $f=\frac{G_{N+1}v_{N+1}c}{N G_{min}}$.

Now, the equilibrium points are given by $\frac{d}{dt}\langle g\rangle=0$. These can be written as those points satisfying
\begin{eqnarray}
\langle g\rangle&=&\frac{\kappa_{P0} e^{\eta_{P} \frac{G_{N+1} v_{N+1}}{G(\langle g\rangle)(1-\Omega_{N+1})}  \frac{c}{N}}}{\kappa_{P0} e^{\eta_{P} \frac{G_{N+1} v_{N+1}}{G(\langle g\rangle)(1-\Omega_{N+1})}  \frac{c}{N}}+\kappa_{D0} e^{\eta_{D} \frac{G_{N+1} v_{N+1}}{G(\langle g\rangle)(1-\Omega_{N+1})}  \frac{c}{N}}}\\
&=&\frac{1}{
1+s e^{\frac{f_0 v}{1+\chi \langle g\rangle}}}
\end{eqnarray}
where we called $f_0=\frac{(\eta_D-\eta_P)c G_{N+1} }{N G_{min}}=(\eta_D-\eta_P)f/v$ and $s=\frac{\kappa_{D0}}{\kappa_{P0}}$.
Now, assuming that $c\sim 1$, $\eta_D-\eta_{P}\approx -10$ $V^{-1}$, $s\approx 5000$. The conductance on the generator is assumed to be negligible, approximately $0.1$ Siemens, while $G_{min}\approx 10^{-3}$ Siemens,  $\chi\approx 10^2$ while $\Omega_{N+1}\approx 1/2$. Thus, $f_0\approx-\frac{2\cdot 10^3}{N} V^{-1}$. See Fig. \ref{fig:mft}.

\begin{figure}
    \centering
    \includegraphics[scale=0.5]{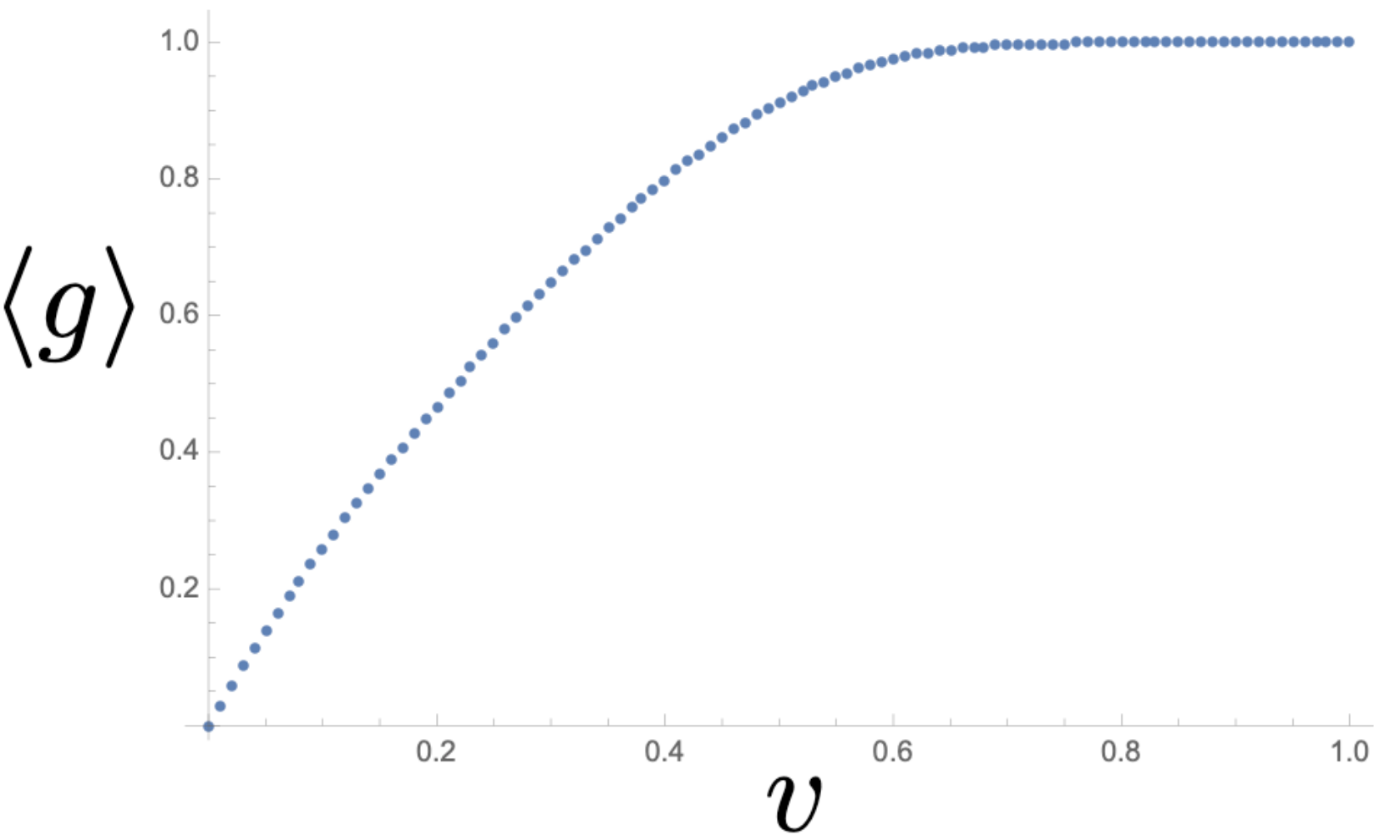}
    \caption{Behavior of $\langle g\rangle$ (y-axis) as a function $v$ (x-axis).}
    \label{fig:mft}
\end{figure}

\subsection{Exact potential}
It is interesting that, using the following exact integrals:
\begin{eqnarray}
\int dx\ e^{\frac{c v}{\chi  x+1}}&=&\frac{(\chi  x+1) e^{\frac{c v}{\chi  x+1}}}{\chi }-\frac{c v \text{Ei}\left(\frac{c v}{x \chi +1}\right)}{\chi }\\
\int dx\  x e^{\frac{d v}{\chi  x+1}}&=&\frac{d v (2-d v) \text{Ei}\left(\frac{d v}{x \chi +1}\right)+\chi  x e^{\frac{d v}{\chi  x+1}} (d v+\chi  x)}{2 \chi
   ^2}+\frac{(d v-1) e^{\frac{d v}{\chi  x+1}}}{2 \chi ^2}
\end{eqnarray}
we can obtain the exact formulation of the potential via these integrals. Let us however focus on the linear regime, and describe the effective potential in detail.

In order to understand the dynamics, let us focus on eqn. (\ref{eq:mftd}), written in order to make the voltage explicit:

\begin{eqnarray}
\frac{d}{dt}\langle g\rangle=\kappa_{P0} e^{\frac{f_{P}v }{1+\chi \langle g\rangle}}  -\Big(\kappa_{P0} e^{\frac{f_{P}v }{1+\chi \langle g\rangle}}+\kappa_{D0} e^{ \frac{f_{D}v }{1+\chi \langle g\rangle}}\Big)\langle g\rangle=-\partial_{\langle g\rangle}V(\langle g\rangle).
\end{eqnarray}

Let us introduce the function
\begin{eqnarray}
\frac{d\langle g\rangle}{dt}&\approx &\kappa_{P0} (1+\frac{f_{P}v }{1+\chi \langle g\rangle})  -\Big(\kappa_{P0} (1+\frac{f_{P}v }{1+\chi \langle g\rangle})+\kappa_{D0} (1+ \frac{f_{D}v }{1+\chi \langle g\rangle})\Big)\langle g\rangle\nonumber \\
&=&\kappa_{P0}+\frac{\kappa_{P0} f_{P}v}{1+\chi \langle g\rangle}-(\kappa_{P0}+\kappa_{D0})\langle g\rangle-\big(\kappa_{P0} f_P+\kappa_{D0}f_D\big) \frac{v \langle g\rangle}{1+\chi \langle g\rangle} \\
&=&-\partial_{\langle g\rangle}V(\langle g\rangle).
\end{eqnarray}
where, writing $\tilde t=\kappa_{P0} t$, we have the adimensional potential
\begin{eqnarray}
\tilde V(\langle g\rangle)=\frac{a \langle g\rangle^2}{2}-v \log (1+\chi  \langle g\rangle) \left(\frac{b}{\chi ^2}+\frac{f_p}{\chi }\right)+\langle g\rangle \left(\frac{b v}{\chi }-1\right)
\end{eqnarray}
where
\begin{eqnarray}
a&=&\frac{\kappa _{\text{D0}}+\kappa _{\text{P0}}}{\kappa _{\text{P0}}}\\
b&=&\frac{f_D \kappa _{\text{D0}}+f_P \kappa _{\text{P0}}}{\kappa _{\text{P0}}}
\end{eqnarray}
and which we can rewrite compactly as
\begin{eqnarray}
\tilde V(v,\langle g\rangle)=a_1(v,\chi) \langle g\rangle+\frac{a_2}{2} \langle g\rangle^2- a_l(\chi) v \log(1+\chi \langle g\rangle)
\end{eqnarray}
This shows that the type of switching is due to a logarithmic potential too.

\section{Gapped nanowires and nanoparticles}

The bulk of the conductance is determined by the quantum tunneling between the hillock and the junctions. The conductance is given by
\begin{eqnarray}
G=\alpha e^{-\beta(D-z)}=\alpha e^{-\beta D(1-z/D)},
\end{eqnarray}
with 
\begin{eqnarray}
\frac{dz}{dt}=r\mu\frac{V}{D-z}-\kappa z
\end{eqnarray}
For $V>V_c$, the junction grows up to $z=D$. Now consider $g=z/D$. We can rewrite the second equation as
\begin{eqnarray}
\frac{dg}{dt}&=&\frac{r\mu}{D^2}\frac{V}{1-g}-\kappa g\\
G(g)&=&\alpha e^{-\beta D(1-g)},
\end{eqnarray}
with $g\in[0,1]$. Note that we can write
\begin{equation}
    G(g)\in[G_{min},G_{max}],
\end{equation}
with $G_{min}=\alpha e^{-\beta D}$ and $G_{max}=\alpha$. We can then rewrite $G=G_{min}(1+\chi f(g))$ with $\chi=\frac{G_{max}-G_{min}}{G_{min}}=\frac{\alpha -\alpha e^{-\beta D}}{\alpha e^{-\beta D}}=e^{\beta D}-1$ and $f(g)=e^{\beta D g}-1$.

We can insert this conductance equation in eqn.(\ref{eq:vapp2}) and obtain
\begin{eqnarray}
 \Delta \vec v&=&\frac{1}{\alpha}G_{N+1} {\tilde B}^t(\tilde B  e^{-\beta D(I-{\bm g})}\tilde {B}^t)^{-1} \tilde B \vec v_s \nonumber \\
  &=&\frac{e^{\beta D}}{\alpha}G_{N+1} {\tilde B}^t(\tilde B  e^{\beta D{\bm g}}\tilde {B}^t)^{-1} \tilde B \vec v_s\\
  &=&\frac{G_{N+1} }{G_{min}} {\tilde B}^t(\tilde B  \tilde {B}^t+\chi \tilde B  {\bm f}({\bm \tilde{ g}})\tilde {B}^t )^{-1} \tilde B \vec v_s.
 \label{eq:vapp3}
\end{eqnarray}
where similarly to what we had before, $g_{N+1}$ is such that $G(g_{N+1})=\alpha e^{-\beta D(1-g_{N+1})}=G_{N+1}$ from which we get
\begin{eqnarray}
 g_{N+1}=-\frac{1}{\beta D}\log \Big(\frac{e^{\beta D}}{\alpha}G_{N+1}\Big).
\end{eqnarray}
In eqn. (\ref{eq:vapp3}), both ${\bm f}({\bm g})$ and $ G_{N+1}/(\alpha e^{-\beta D})$ are adimensional.
As in eqn. (\ref{eq:split}), we write
\begin{eqnarray}
 \Delta \vec v= {\tilde B}^t(\tilde B\tilde {B}^t+\chi \tilde B{\bm f}({\bm g})\tilde {B}^t)^{-1} \tilde B \vec v_s.
 \label{eq:split2}
\end{eqnarray}
 We can now write
\begin{eqnarray}
 \Delta \vec v= \frac{G_{N+1} }{G_{min}} (I+\chi \tilde \Omega_{\tilde B}{\bm f}({\bm g}))^{-1} \tilde \Omega_{\tilde B} \vec v_s.
 \label{eq:split3}
\end{eqnarray}
Taking the first, $N$ components of the equation above, $\Delta \vec v_N$, we can insert the expression above, now, in the network equation for the conductances. We have
\begin{eqnarray}
\frac{d\vec g}{dt}&=&\frac{r\mu}{D^2}(I-{\bm g})^{-1}\Delta \vec v_N-\kappa \vec g\\
{\bm G}({\bm g})&=&\alpha e^{-\beta D(I-{\bm g})}.
\end{eqnarray}
As we did before, the issue now is how to block invert $(I+\chi \tilde \Omega_{\tilde B} {\bm f}( {\bm g}))^{-1}$. Above, we can imagine that ${\bm f}( {\bm g})$ is a general function.
Let us use
\begin{eqnarray}
(I+\chi \tilde \Omega_{\tilde B}{\bm f}({\bm g}))^{-1}&=&\begin{pmatrix}
I+\chi \tilde \Omega {\bm f}(\tilde {\bm{g}}) & \chi f(g_{N+1}) \vec \Omega \\
\chi ( f( \bm{g}) \vec \Omega)^t & 1+\chi f(g_{N+1}) \Omega_{N+1}
\end{pmatrix}^{-1}
\end{eqnarray}

We now use the matrix block inverse identity
\begin{eqnarray}
\begin{pmatrix}
A & B \\
C & D
\end{pmatrix}^{-1}=\begin{pmatrix}
Q_{11} & \vec Q_{12}\\
\vec Q_{21}^t & Q_{22}
\end{pmatrix}=\begin{pmatrix}
A^{-1}+A^{-1} B(D-C A^{-1} B)^{-1} C A^{-1} & -A^{-1} B (D-CA^{-1 }B)^{-1} \\
-(D-C A^{-1} B)^{-1} C A^{-1} & (D-CA^{-1} B)^{-1}
\end{pmatrix}
\end{eqnarray}
We have $A=I+\chi \tilde \Omega \tilde {\bm{g}}$, which we stress is a $N\times N$ matrix.
Let us now focus on $q_0=D-CA^{-1} B$.
This quantity is a scalar, given by
\begin{eqnarray}
q_0&=&1+\chi g_{N+1}( \Omega_{N+1}-\chi \vec \Omega^{t} {\bm f}(\tilde{\bm{g}}) A^{-1} \vec \Omega).
\label{eq:q0}
\end{eqnarray}
We define the rank-1 matrix $$r_0=\chi^2 g_{N+1} (\vec \Omega) \otimes ({\bm f}\big({\tilde {\bm{ g}}})\vec \Omega\big)^t.$$
First, we have $$Q_{22}=q_0^{-1}.$$
Also, after a quick calculation we can show that
\begin{eqnarray}
Q_{11}=(I+\chi \tilde \Omega {\bm f}(\tilde{\bm{g}}))^{-1} +q_0^{-1} (I+\chi \tilde \Omega {\bm f}(\tilde{\bm{g}}))^{-1}r_0 (I+\chi \tilde \Omega {\bm f}(\tilde{\bm{g}}))^{-1}
\end{eqnarray}

and we get
\begin{eqnarray}
\vec {Q}_{12}=-\chi\frac{g_{N+1}}{q_0} (I+\chi \tilde \Omega {\bm f}(\tilde{\bm{g}}))^{-1} \vec \Omega,
\end{eqnarray}
while
\begin{eqnarray}
\vec Q_{21}^t=-\chi \frac{1}{q_0} \vec \Omega^t{\bm f}(\tilde{\bm{g}})(I+\chi \tilde \Omega {\bm f}(\tilde{\bm{g}}))^{-1}  .
\end{eqnarray}
Let us now focus on $\Omega \vec v_s$. We write
\begin{eqnarray}
\vec v_s=\begin{pmatrix}
\tilde v\\
v_{N+1}
\end{pmatrix}
\end{eqnarray}
thus we have
\begin{eqnarray}
\Omega \vec v_s=\begin{pmatrix}
\tilde \Omega & \vec \Omega\\
\vec \Omega^t & \Omega_{N+1}
\end{pmatrix}\begin{pmatrix}
\tilde v\\
v_{N+1}
\end{pmatrix}=\begin{pmatrix}
\tilde \Omega \tilde v+v_{N+1} \vec \Omega\\
\Omega_{N+1} v_{N+1}+\vec \Omega^t \tilde v
\end{pmatrix}=\begin{pmatrix}
\vec v_a\\
v_b
\end{pmatrix}
\end{eqnarray}
and thus we get
\begin{eqnarray}
(I+\chi \Omega {\bm f}({\bm{g}}))^{-1} \Omega \vec v_s=\begin{pmatrix}
Q_{11} & \vec Q_{12}\\
\vec Q_{21}^t & Q_{22}
\end{pmatrix}\begin{pmatrix}
\vec v_a\\
v_b
\end{pmatrix}=\begin{pmatrix}
R_{1}\\
R_{2} 
\end{pmatrix}
\end{eqnarray}
where
\begin{eqnarray}
R_{1}&=&Q_{11} \vec v_a+ \vec Q_{12} v_b= Q_{11}(\tilde \Omega \tilde v+ v_{N+1} \vec \Omega)+(\Omega_{N+1} v_{N+1}+ \vec \Omega^t \tilde v_a)\vec Q_{12}\\
R_{2} &=& \vec Q_{21}^t \vec v_a+Q_{22} v_b=\vec Q_{21}^t (\tilde \Omega \tilde v+ v_{N+1} \vec \Omega)+(\Omega_{N+1}v_{N+1}+\vec \Omega^t \tilde v)Q_{22}
\end{eqnarray}
If we now assume that $\tilde v=0$, we get
\begin{eqnarray}
R_{1}&=& v_{N+1}(Q_{11} \vec \Omega+\Omega_{N+1} \vec Q_{12})\\
R_{2} &=&   v_{N+1} (\vec Q_{21}^t\vec \Omega+\Omega_{N+1}Q_{22})
\end{eqnarray}
and thus we get
\begin{eqnarray}
\Delta\vec v_{all}&= &\frac{G_{N+1}}{G_{min}}\begin{pmatrix}
R_{1}\\
R_{2} 
\end{pmatrix}\nonumber \\
&=&v_{N+1}\frac{G_{N+1}}{G_{min}}\begin{pmatrix}
Q_{11} \vec \Omega+\Omega_{N+1} \vec Q_{12}\\
\vec Q_{21}^t\vec \Omega+\Omega_{N+1}Q_{22}
\end{pmatrix}\\
&=&\frac{G_{N+1}v_{N+1}}{G_{min}}\begin{pmatrix}
&&\Big((I+\chi \tilde \Omega {\bm f}({ \tilde{\bm g}})^{-1} +q_0^{-1} (I+\chi \tilde \Omega {\bm f}({ \tilde{\bm g}}))^{-1}f_0 (I+\chi \tilde \Omega {\bm f}({ \tilde{\bm g}}))^{-1}\Big)\vec \Omega\nonumber \\
&&(ctd)\ \ \ \ \ \ \ \ \ -g_{N+1}\chi q_0^{-1}\Omega_{N+1}(I+\chi \tilde \Omega {\bm f}({ \tilde{\bm g}}))^{-1} \vec \Omega\nonumber  \\
&&-\frac{1}{q_0}(\chi\vec \Omega^t {\bm f}({ \tilde{\bm g}})(I+\chi \tilde\Omega{\bm f}({ \tilde{\bm g}}))^{-1}\vec \Omega-\Omega_{N+1})
\end{pmatrix}\\
&=& \frac{G_{N+1}v_{N+1}}{G_{min}}\begin{pmatrix}
\Big(I +\frac{ (I+\chi \tilde \Omega {\bm f}({ \tilde{\bm g}}))^{-1}r_0-\chi\Omega_{N+1} g_{N+1} I }{1+\chi g_{N+1}( \Omega_{N+1}- \chi\vec \Omega^{t}{\bm f}({ \tilde{\bm g}}) A^{-1} \vec \Omega)} \Big)(I+\chi \tilde \Omega {\bm f}({ \tilde{\bm g}}))^{-1}\vec \Omega\\
\frac{\rho}{q_0}
\end{pmatrix}
\end{eqnarray}
where we have 
called
\begin{eqnarray}
\rho= \Omega_{N+1}- \chi\vec \Omega^{t}{\bm f}({ \tilde{\bm g}}) (I+\chi \tilde \Omega {\bm f}({ \tilde{\bm g}}))^{-1} \vec \Omega
\label{eq:etasm2}
\end{eqnarray}

Note that if $G_{N+1}\rightarrow 0$, we have $g_{N+1}\rightarrow -\frac{1}{\chi}$.
The formula above does not have any approximations.

The voltage drop on the devices is the vector of internal voltage drops, given by 
\begin{eqnarray}
\Delta \vec v_{int}&=&\frac{G_{N+1}v_{N+1}}{G_{min}}
\Big(I +\frac{ (I+\chi \tilde \Omega {\bm f}({ \tilde{\bm g}}))^{-1}r_0-\chi\Omega_{N+1} g_{N+1} I }{1+\chi g_{N+1}( \Omega_{N+1}- \chi\vec \Omega^{t}{\bm f}({ \tilde{\bm g}}) (I+\chi \tilde \Omega {\bm f}({ \tilde{\bm g}}))^{-1} \vec \Omega)} \Big)(I+\chi \tilde \Omega {\bm f}({ \tilde{\bm g}}))^{-1}\vec \Omega\nonumber \\
&=&\frac{G_{N+1}v_{N+1}}{G_{min}}
\Big(I +\frac{ g_{N+1}\chi^2(I+\chi \tilde \Omega {\bm f}({ \tilde{\bm g}}))^{-1}  (\vec \Omega) \otimes \big({\bm f}({ \tilde{\bm g}})\vec \Omega\big)^t-\chi\Omega_{N+1} g_{N+1} I }{1+\chi g_{N+1}( \Omega_{N+1}- \chi\vec \Omega^{t}{\bm f}({ \tilde{\bm g}}) (I+\chi \tilde \Omega {\bm f}({ \tilde{\bm g}}))^{-1} \vec \Omega)} \Big)(I+\chi \tilde \Omega {\bm f}({ \tilde{\bm g}}))^{-1}\vec \Omega\nonumber \\
&=&\frac{G_{N+1}v_{N+1}}{G_{min}}
\Big(I +g_{N+1}\chi\frac{ \chi \vec \Omega^t{\tilde {\bm{ g}}}(I+\chi \tilde \Omega {\bm f}({ \tilde{\bm g}}))^{-1}  \vec \Omega -\Omega_{N+1}   }{1+\chi g_{N+1}( \Omega_{N+1}- \chi\vec \Omega^{t}{\bm f}({ \tilde{\bm g}}) (I+\chi \tilde \Omega {\bm f}({ \tilde{\bm g}}))^{-1} \vec \Omega)} \Big)(I+\chi \tilde \Omega {\bm f}({ \tilde{\bm g}}))^{-1}\vec \Omega\nonumber \\
\end{eqnarray}
and using the fact that $g_{N+1}\approx -1/\chi$, we have

\begin{eqnarray}
\Delta \vec v_{int}
&=&\frac{G_{N+1}v_{N+1}}{G_{min}}
\Big(1 +\frac{ \rho}{1-\rho} \Big)(I+\chi \tilde \Omega {\bm f}({ \tilde{\bm g}}))^{-1}\vec \Omega\nonumber \\
&=&\frac{1}{1-\rho}\frac{G_{N+1}v_{N+1}}{G_{min}}
(I+\chi \tilde \Omega {\bm f}({ \tilde{\bm g}}))^{-1}\vec \Omega
\label{eq:vexact}
\end{eqnarray}

\subsection{Effective conductance}
Let us now use the equations above to obtain results about the effective conductance of the whole device, where we can use eqn. (\ref{eq:effcond}), which we recall is
\begin{eqnarray}
G_{eff}=\frac{j_{N+1}}{\Delta v_{N+1}}.
\end{eqnarray}
We can then note that, from the equations above, we have
\begin{eqnarray}
\Delta v_{N+1}=\frac{G_{N+1} v_{N+1}}{G_{min}} \frac{\rho}{q_0}.
\end{eqnarray}
We then have explicit expressions for $q_0$ and $\rho$, eqn. (\ref{eq:q0}) and eqn. (\ref{eq:etasm}) respectively.
Replacing, we have
\begin{eqnarray}
G_{eff}&=&\frac{j_{N+1}}{\frac{G_{N+1}v_{N+1}}{G_{min}}\frac{\eta}{q_0} }= G_{min}\frac{j_{N+1}}{G_{N+1}v_{N+1} }\frac{q_0}{\eta}\\
&=&G_{min}\frac{j_{N+1}}{G_{N+1}v_{N+1} }\frac{1+\chi g_{N+1}( \Omega_{N+1}-\chi \vec \Omega^{t}{\bm f}({ \tilde{\bm g}}) (I+\chi \tilde \Omega {\bm f}({ \tilde{\bm g}}))^{-1} \vec \Omega)}{\Omega_{N+1}- \chi\vec \Omega^{t}{\bm f}({ \tilde{\bm g}}) (I+\chi \tilde \Omega {\bm f}({ \tilde{\bm g}}))^{-1} \vec \Omega}\\
&=&G_{min}\frac{j_{N+1}}{G_{N+1}v_{N+1} }\frac{1+\chi g_{N+1} \eta}{\eta}.
\end{eqnarray}

We note that $G_{n+1}{v_{N+1}}=j_{N+1}$, and thus the expression simplifies to
\begin{eqnarray}
G_{eff}&=&G_{min}\frac{1+\chi g_{N+1}\rho}{\rho}=G_{min}\frac{1+\chi g_{N+1}( \Omega_{N+1}-\chi \vec \Omega^{t}{\bm f}({ \tilde{\bm g}}) (I+\chi \tilde \Omega {\bm f}({ \tilde{\bm g}}))^{-1} \vec \Omega)}{\Omega_{N+1}- \chi\vec \Omega^{t}{\bm f}({ \tilde{\bm g}}) (I+\chi \tilde \Omega {\bm f}({ \tilde{\bm g}}))^{-1} \vec \Omega}
\end{eqnarray}
which is an exact expression of the effective conductance of the whole device. Now, in the limit $G_{N+1}\rightarrow 0$, we have $g_{N+1}\rightarrow-\frac{1}{\chi}$. 

\subsection{Matrix inverse approach and mean field theory voltage drops}
As mentioned, the issue is the matrix inverse given by
\begin{eqnarray}
(I+\chi \tilde \Omega {\bm f}({ \tilde{\bm g}}))^{-1}.
\end{eqnarray}
We then wonder if we could perform the approximation
\begin{eqnarray}
(I+\chi \tilde \Omega {\bm f}({ \tilde{\bm g}}))^{-1}\approx (I+\chi \tilde \Omega f(\langle g\rangle))^{-1}
\end{eqnarray}
where $\langle g\rangle$ is the minimum of a matrix norm of the form
\begin{eqnarray}
\langle g\rangle&=&\text{argmin}_{\langle g\rangle} \text{Tr}\big((I+\chi\tilde \Omega {\bm f}({ \tilde{\bm g}}))- (I+\chi \tilde \Omega f(\langle g\rangle)) \big)^2 \\
&=&\chi^2\text{argmin}_{\langle g\rangle} \text{Tr}\big(\tilde \Omega {\bm f}({ \tilde{\bm g}})-  \tilde \Omega f(\langle g\rangle) \big)^2\\
&=& \chi^2\text{argmin}_{\langle g\rangle} \text{Tr}\big((\tilde \Omega {\bm f}({ \tilde{\bm g}}))^2+  (\tilde \Omega f(\langle g\rangle))^2 -2f(\langle g\rangle)\tilde \Omega^2{\bm f}({ \tilde{\bm g}})\big)
\end{eqnarray}
We can derive with respect to $\langle g\rangle$ the Frobenius norm, obtaining the maximization
\begin{eqnarray}
0&=&\partial_{\langle g\rangle}\chi^2\text{argmin}_{\langle g\rangle} \text{Tr}\big((\tilde \Omega {\bm f}({ \tilde{\bm g}}))^2+  (\tilde \Omega f(\langle g\rangle))^2 -2f(\langle g\rangle)\tilde \Omega^2{\bm f}({ \tilde{\bm g}})\big)\\
&=&2\chi^2f'(\langle g\rangle)\text{argmin}_{\langle g\rangle} \text{Tr}\big(\tilde \Omega^2f(\langle g\rangle) -\tilde \Omega^2{\bm f}({ \tilde{\bm g}})\big)\\
\end{eqnarray}
from which we get the value
\begin{eqnarray}
\langle g\rangle=f^{-1}\Big(\frac{\text{Tr}(\tilde \Omega^2{\bm f}({ \tilde{\bm g}}))}{\text{Tr}(\tilde \Omega^2)}\Big)
\end{eqnarray}
Note that $\tilde \Omega$ is not a projector, and thus $\tilde \Omega^2\neq \tilde \Omega$.
It follows that we have the norm-2 approximation to the inverse, given by
\begin{eqnarray}
(I+\chi \tilde \Omega {\bm f}({ \tilde{\bm g}}))^{-1}\approx (I+\chi \tilde \Omega \frac{\text{Tr}(\tilde \Omega^2{\bm f}({ \tilde{\bm g}}))}{\text{Tr}(\tilde \Omega^2)})^{-1}\equiv (I+\chi \tilde \Omega f(\langle g\rangle))^{-1},
\end{eqnarray}
which we use next. Note that the approximation is consistent, e.g. that if $\bm{g}=\langle g\rangle I$. Thus, the more homogeneous the memristors are the more precise the mean-field theory becomes.

We can use the expression above for a mean-field theory. If we use $\tilde{\bm{g}}=\langle g\rangle I$, where $\langle g\rangle$ has to be determined, and then we get
\begin{eqnarray}
\Delta \vec v\approx \frac{G_{N+1} v_{N+1}}{G_{min}(1-\rho)} (I+\chi f(\langle g\rangle) \tilde \Omega)^{-1}\vec \Omega=\frac{G_{N+1} v_{N+1}}{G_{min}(1-\rho)} \frac{1}{1+\chi f(\langle g\rangle)(1-\Omega_{N+1})}\vec \Omega
\label{eq:DeltaVBrown}
\end{eqnarray}
where $\langle g\rangle$, as in the case of the linear memristor, it is chosen to minimize the matrix inverse deviation,
and for $\rho$ we have
\begin{eqnarray}
\rho&\approx& \Omega_{N+1}-\chi \frac{f(\langle g\rangle)}{1+\chi f(\langle g\rangle)(1-\Omega_{N+1})}\vec \Omega^t \vec \Omega\\
&=&\Omega_{N+1}-\chi \frac{f(\langle g\rangle)}{1+\chi f(\langle g\rangle)(1-\Omega_{N+1})}\Omega_{N+1}(1-\Omega_{N+1})\\
&=&\Omega_{N+1}\Big(1-\frac{f(\langle g\rangle)\chi (1-\Omega_{N+1})}{1+f(\langle g\rangle)\chi (1-\Omega_{N+1})}\Big)=\frac{\Omega_{N+1}}{{1+f(\langle g\rangle)\chi (1-\Omega_{N+1})}}
\end{eqnarray}
and thus, following the proof
\begin{eqnarray}
1-\rho&=&\frac{{1+f(\langle g\rangle)\chi (1-\Omega_{N+1})}}{{1+f(\langle g\rangle)\chi (1-\Omega_{N+1})}}-\frac{\Omega_{N+1}}{{1+f(\langle g\rangle)\chi (1-\Omega_{N+1})}}\\
&=&\frac{1+f(\langle g\rangle) \chi-\Omega_{N+1}(1+f(\langle g\rangle) \chi)}{{1+f(\langle g\rangle)\chi (1-\Omega_{N+1})}}=(1-\Omega_{N+1})\frac{1+f(\langle g\rangle)\chi}{1+f(\langle g\rangle)\chi(1-\Omega_{N+1})}
\end{eqnarray}
from which we then get
\begin{eqnarray}
\Delta \vec v\approx \Delta \vec v_{mft}=\frac{G_{N+1} v_{N+1}}{G_{min}(1-\Omega_{N+1})} \frac{1}{1+f(\langle g\rangle) \chi}\vec \Omega.
\label{eq:vmft}
\end{eqnarray}
At this point, we can also write a formula for the effective conductance in terms of the effective conductance, given by

\begin{eqnarray}
G_{eff}&=&G_{min}\frac{1-\rho}{\rho}=G_{min}\frac{(1-\Omega_{N+1})\frac{1+f(\langle g\rangle)\chi}{1+f(\langle g\rangle)\chi(1-\Omega_{N+1})}}{\frac{\Omega_{N+1}}{{1+f(\langle g\rangle)\chi (1-\Omega_{N+1})}}}=\frac{1-\Omega_{N+1}}{\Omega_{N+1}}(1+\chi f(\langle g\rangle))\\
&=&\frac{1-\Omega_{N+1}}{\Omega_{N+1}}G(\langle g\rangle)
\end{eqnarray}
which is the equation for a global memristor.

\subsection{Mean field for nanoparticles}
We then obtain that, starting from the model for the nanowires given by
\begin{eqnarray}
  \frac{dx}{dt}=\mu\frac{V}{D-x}-\kappa x, G(x)=\alpha e^{-\beta (D-x)}.
\end{eqnarray}
If we write $g=x/D$, we have the equations
\begin{eqnarray}
  \frac{dg}{dt}=\mu\frac{V}{D^2(1-g)}-\kappa g, G(x)=\alpha e^{-\beta/D (1-g)},
\end{eqnarray}
with $0\leq g \leq 1$.
For a network, the equations become vectorial. We have
\begin{eqnarray}
\frac{d\vec g}{dt}&=&\frac{r\mu}{D^2}(I-{\bm g})^{-1}\Delta \vec v_N-\kappa \vec g\\
{\bm G}({\bm g})&=&\alpha e^{-\beta D(I-{\bm g})}.
\end{eqnarray}
We can now replace eqn. (\ref{eq:DeltaVBrown}), and obtain in the first equation
\begin{eqnarray}
\frac{d\vec g}{dt}&\approx&\frac{r\mu}{D^2}(I-{\bm g})^{-1}\frac{G_{N+1} v_{N+1}}{G_{min}(1-\eta)} \frac{1}{1+\chi f(\langle g\rangle)(1-\Omega_{N+1})}\vec \Omega-\kappa \vec g\\
{\bm G}({\bm g})&=&\alpha e^{-\beta D(I-{\bm g})}.
\end{eqnarray}
with $f(\langle g\rangle)=e^{\beta D \langle g\rangle}-1$ and $\langle g\rangle=f^{-1}\Big(\frac{1}{\text{Tr}(\tilde \Omega^2)} \text{Tr}\big(\tilde \Omega^2 f({\bm {\tilde g}})\big) \Big).$

Then, the effective mean field can be obtained by imposing
$\vec g=\langle g\rangle \vec 1$, and then we end up with the effective equations
\begin{eqnarray}
\frac{d}{dt}\langle g\rangle&=&\frac{r\mu}{D^2}(I-\langle g\rangle)^{-1}\frac{G_{N+1} v_{N+1}}{G_{min}(1-\eta)} \frac{1}{1+\chi f(\langle g\rangle)(1-\Omega_{N+1})}\langle \vec \Omega\rangle-\kappa \langle g\rangle\\
G_{eff}(\langle g\rangle) &=&\frac{1-\Omega_{N+1}}{\Omega_{N+1}}G(\langle g\rangle)\equiv \rho G(\langle g\rangle)
\end{eqnarray}
where $\langle \vec \Omega\rangle=\frac{1}{N}\sum_{i} (\vec \Omega)_i$.

The equation above can be rewritten, in terms of the effective parameters, as
\begin{eqnarray}
\frac{d}{dt}\langle g\rangle&=& \frac{q_{eff}}{\big(1-\langle g\rangle\big)\big(1+\chi_{eff} f(\langle g\rangle)\big)}-\kappa_{eff} \langle g\rangle\\
G_{eff}(\langle g\rangle) &=&G_{min}^{eff}\ (1+\chi_{eff}f(\langle g\rangle))
\end{eqnarray}
with $f(x)=e^{-a(1-x)}-1$. This is the model studied in the main text.

\fi
\end{document}